\documentclass{aa}

\usepackage[varg]{txfonts}
\usepackage{graphicx}
\usepackage{array, booktabs, makecell}
\usepackage{siunitx}
\usepackage[T1]{fontenc}
\usepackage{newtxtext,newtxmath}
\usepackage[version=4]{mhchem}
\usepackage{textcomp, gensymb}
\usepackage{xcolor}
\usepackage{amsmath}
\usepackage{amsfonts}
\usepackage{comment}
\usepackage{multicol}
\usepackage[utf8]{inputenc}
\usepackage[english]{babel}
\usepackage{natbib}
\bibpunct{(}{)}{;}{a}{}{,}
\usepackage[hidelinks,colorlinks=true,linkcolor=blue,citecolor=blue]{hyperref}

\begin{document}

\title{
Exploring the disk-jet connection in NGC\,315} 

\author{L.\ Ricci \inst{1}, 
        B.\ Boccardi \inst{1},
        E.\ Nokhrina \inst{2},
        M.\ Perucho \inst{3,4},
        N.\ MacDonald \inst{1},
        G.\ Mattia \inst{5},
        P.\ Grandi \inst{6},
        E.\ Madika \inst{1},
        T.\ P.\ Krichbaum \inst{1}, and
        J.\ A.\ Zensus \inst{1}\
     } 

\institute{
\inst{1} Max-Planck-Institut f\"{u}r Radioastronomie, Auf dem H\"{u}gel 69, D-53121 Bonn, Germany \\
\inst{2} Moscow Institute of Physics and Technology, Dolgoprudny, Institutsky per., 9, Moscow region, 141700, Russia \\
\inst{3} Departament d’Astronomia i Astrofísica, Universitat de València, C/ Dr. Moliner, 50, 46100, Burjassot, València, Spain \\
\inst{4} Observatori Astronòmic, Universitat de València, C/ Catedràtic José Beltrán 2, 46980, Paterna, València, Spain \\
\inst{5} Max Planck Institute f\"{u}r Astronomie, Königstuhl 17, 69117, Heidelberg, Germany \\
\inst{6} INAF – Osservatorio di Astrofisica e Scienza dello Spazio di Bologna, Via Gobetti 101, I-40129 Bologna, Italy \\
}

\date{}

  \abstract
   {}
{Hot accretion flows are thought to be able to power the relativistic jets observed in Active Galactic Nuclei. They can present themselves as SANE (Standard And Normal Evolution) disks or MAD (Magnetically Arrested Disks), two states implying profound differences in the physical properties of the disks themselves and of the outflows they produce.}
{In this paper we use a multi-frequency and multi-epoch data set to study the giant radio galaxy NGC\,315, with the goal to explore the properties of its accretion disk and sub-parsec jet. We analyze the source maps with a pixel-based analysis and we use theoretical models to link the observational properties of the jet to the physical state of the accretion disk.}
{We propose that the bulk flow in NGC\,315 accelerates on sub-pc scales, concurrently with the parabolic expansion. We show that this fast acceleration can be theoretically reconciled with a magnetically driven acceleration.
Along the acceleration and collimation zone, we observe an unexpected spectral behavior, with very steep spectral index values $\alpha \sim -1.5$ ($S_\nu \propto \nu^\alpha$) between 22 GHz and 43 GHz. 
Based on the properties of this region, we predict the black hole of NGC\,315 to be fast rotating and the magnetic flux threading the accretion disk to be in excellent agreement with that expected in the case of a MAD.
Using a new formalism based on the core-shift effect, we model the magnetic field downstream a quasi-parabolic accelerating jet and we reconstruct it up to the event horizon radius.
In the MAD scenario, we compare it with the expected magnetic saturation strengths in the disk, finding a good agreement.

}
   {}
\titlerunning{Exploring the disk-jet connection in NGC\,315}
\authorrunning{L. Ricci et al.}
\maketitle

\section{Introduction} \label{sec:introduction}

The disk-jet connection is one of the most compelling and studied topics concerning the physics of Active Galactic Nuclei (AGN),
with the efforts of the community being focused on unveiling the nature of the accretion disks that are capable to launch different categories of jets.
Hot accretion flows are able to produce the powerful relativistic jets we observe in AGN. 
They may present themselves in two different flavours, namely as SANE \citep[Standard And Normal Evolution,][]{Narayan_2012}
disks or MAD \citep[Magnetically Arrested Disk,][]{Bisno_1974, Bisno_1976, Narayan_2003}.
In the first case, the magnetic field is not dynamically important and it only generates effective viscosity, facilitating accretion through the outward transport of the angular momentum.
In the MAD scenario, in contrast, the disk is naturally permeated by strong magnetic fields which are carried by the inflowing matter.
This leads the magnetic field to gradually accumulate in the vicinity of the central black hole.
Once a certain threshold is reached, the magnetic field becomes dynamically important and its strength is high enough to balance the gravitational force of the accreting gas, blocking its inflow and creating a magnetosphere.
In the magnetosphere, the torque exerted by the rotation of the central black hole alters the magnetic field lines, which bend toward the directions perpendicular to the disk plane. Since the gas within the accretion disk is in a plasma state, the charged particles are forced to follow the magnetic field lines, and jets are launched \citep{Blandford_1977}.
The two disk states differ in the magnetic field distribution and strength both in the disk and in the resultant jets.


The most direct way to distinguish between the SANE and MAD state is by observing the polarization and Faraday rotation of the central core and the inner jet region
\citep[][and references therein]{Yuan_2022}.
When the polarization is not seen or is very weak, which is the case for most radio galaxies, we need to rely on alternative methods based on the indirect estimation of the magnetic activity in the disk.
This is possible since the physical properties of the accretion disk are closely related to some observational properties of the jets on VLBI (Very Long Baseline Interferometry) scales, such as the extent of the collimation and acceleration region.
In particular, the jets are initially dominated by the magnetic energy extracted from the accretion disk and propagate as pure Poynting flux. 
While expanding within the external medium, they collimate showing a parabolic shape on scales of $10^3- 10^7 \, R_\mathrm{S}$ \footnote{$R_\mathrm{S}$ is the Schwarzschild radius defined as $R_\mathrm{S} = 2 G M_\mathrm{BH} / c^2$ where $M_\mathrm{BH}$ is the mass of the black hole} \citep{Kovalev_2020,Boccardi_2021}, and accelerate up to relativistic velocities.
The acceleration and collimation processes are expected to be co-spatial in the case of a magnetically-driven cold outflow \citep{Komissarov_2007, Tchekhovskoy_2008, Lyubarsky_2009}. 
The jets are accelerated mainly through the conversion of the initial magnetic energy into kinetic energy of the bulk flow and radiation, and this process is believed to continue until equipartition between the particles energy and the magnetic field is reached \citep{Komissarov_2009}.
Therefore, by examining the properties of the acceleration and collimation in a jet, it is possible to reconstruct, by means of theoretical models, the magnetic field that permeates the accretion disk and explore the physical state of the disk itself.

The properties of such magnetic fields are highly debated.
Estimates based on the assumption of a conical jet in equipartition \citep{Lobanov_1998, Hirotani_2005} result in magnetic field strengths of hundreds of mG at 1 pc from the core 
\citep[e.g.,][]{O'Sullivan_2009}, which implies strengths of $\sim 10^3-10^4 \, \mathrm{G}$ in the surroundings of the black hole, assuming a toroidal-dominated magnetic field evolution ($B \propto \mathsf{z}^{-1}$, where $\mathsf{z}$ denotes the distance from the black hole along the jet). 
However, rotation measure studies have provided evidence of helical fields along the acceleration region
\citep[][and references therein]{Hovatta_2017}, suggesting that the poloidal field component ($B \propto \mathsf{z}^{-2}$ in a conical jet) cannot be neglected when extrapolating the magnetic field strength.
The steeper evolution implies higher field strengths in the core region.
Magnetic fields stronger than $\sim 10^4 \, \mathrm{G}$ are not sustainable by black holes and may point to the existence of exotic objects in the core of the galaxies, such as gravastars or wormholes \citep{Lobanov_2017}.

In this paper, we explore the nearby
\citep[$z = 0.0165$,][]{Trager_2000} giant radio galaxy NGC\,315 through VLBI observations. 
NGC\,315 extends in total up to $\sim 1 \, \mathrm{Mpc}$ and shows a Fanaroff--Riley \citep{Faranoff_Riley} I morphology
\citep[e.g.,][]{Laing_2006}. 
In NGC\,315 the acceleration and collimation region is observed on VLBI scales \citep{Boccardi_2021, Park_2021}, making it a perfect target to investigate the disk-jet connection.

The paper is organized as follows. In Sect.\ \ref{sec:Data_set} we describe the data set and the methods used for the analysis; in Sect.\ \ref{sec:Results} we present our observational results; in Sect.\ \ref{sec:Discussion} we apply different theoretical models to interpret our results, and in Sect.\ \ref{sec:Conclusions} we draw our conclusions. 


We assume a $\Lambda$CDM cosmology with $H_0 = 71 \ \mathrm{h \ km \ s^{-1} \ Mpc^{-1}}$, $\Omega_M = 0.27$, $\Omega_\Lambda = 0.73$ \citep{Komatsu}.
At the redshift of NGC\,315, the luminosity distance is $D_L = 70.6 \, \mathrm{Mpc}$ and 1 mas corresponds to 0.331 pc.

\section{Data set and image analysis} \label{sec:Data_set}

\begin{figure}[t]
  \resizebox{\hsize}{!}{\includegraphics{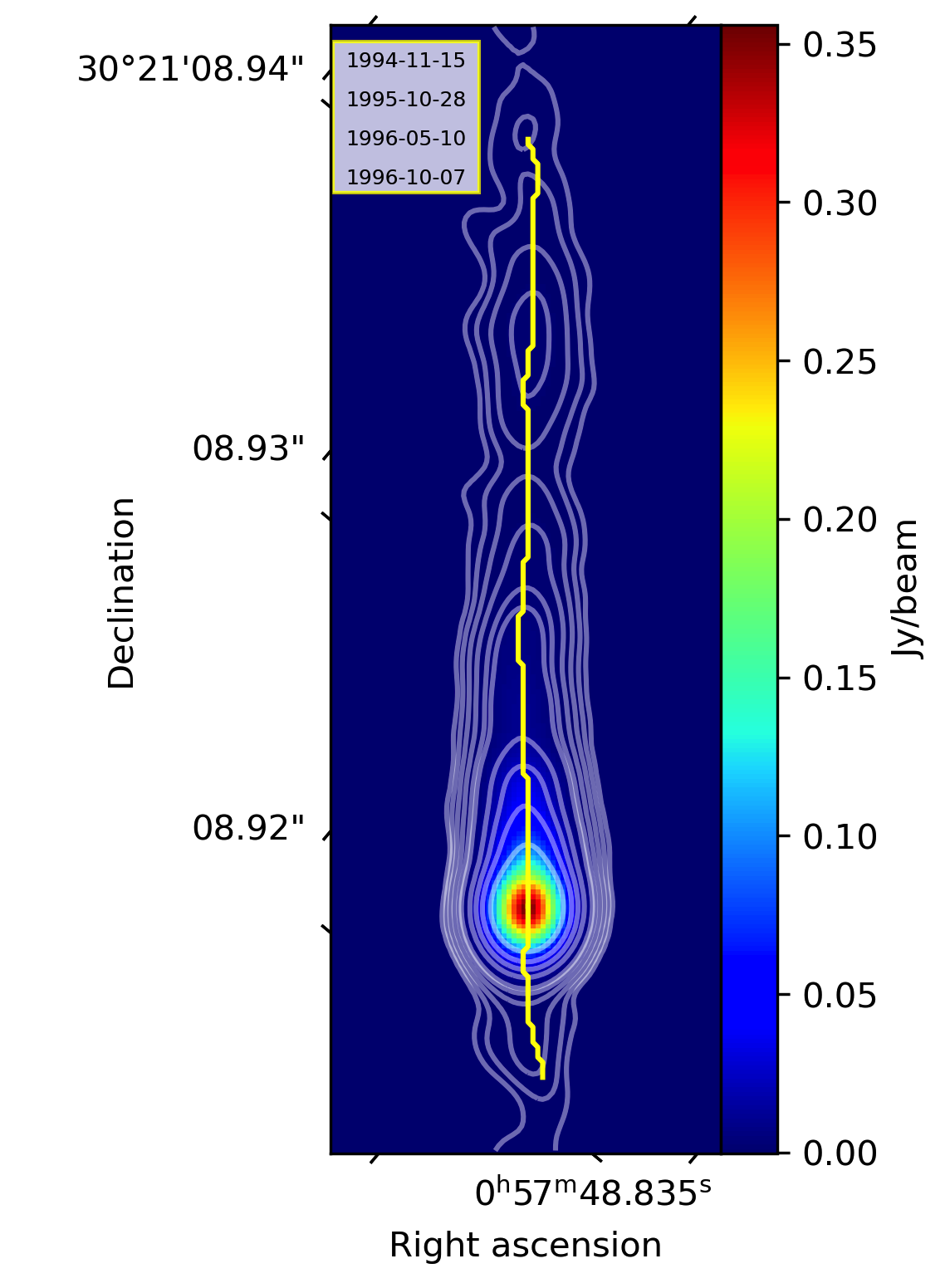}}
  \caption{5 GHz stacked VLBI image obtained using the four observations reported in figure. The contour levels are set at [0.001, 0.002, 0.004, 0.008, 0.012, 0.034, 0.068, 0.126, 0.254] $\times$ the peak brightness of 0.356 Jy/beam. The jet ridgeline is highlighted by the yellow line, and is found to be remarkably straight.}
  \label{fig:ridgeline}
\end{figure}

In this article, we consider the multi-frequency and multi-epoch VLBI data set presented by \citet{Boccardi_2021} (hereafter B21), which consists of eighteen observations in the frequency range 1 GHz-86 GHz. In particular, since we are interested in utilising the symmetry properties of the approaching and receding jets to derive fundamental jet parameters, we focus on those epochs where both the jet and the counter-jet were detected.
The main properties of the maps are summarized in Table \ref{tab:beam}. Together with the B21 observations, we analyzed one image from 2019 at 15 GHz from the MOJAVE program \citep{Lister_2018}, since a small counter-jet is detected.
The images used in our analysis are shown in Appendix \ref{app:MOJAVE_image}.
In addition, we analyze stacked images at 5 GHz and 15 GHz.
The 5 GHz stacked image is created by considering four different observations in the period 1994-1996 (see Table 1, B21), while the 15 GHz stacked image is obtained from the MOJAVE archive, and was already analyzed by \citet{Pushkarev_2017, Kovalev_2020}.
The 5 GHz stacked image with the overlaid jet ridgeline is shown in Fig.\ \ref{fig:ridgeline}.

\begin{table}[]
\caption{Characteristics of the CLEAN maps considered in this article.}
\begin{tabular}{cccc}
\hline
\multicolumn{1}{c|}{\begin{tabular}[c]{@{}c@{}}Freq.\\ {[}GHz{]}\end{tabular}} & \multicolumn{1}{c|}{Date} & \multicolumn{1}{c|}{\begin{tabular}[c]{@{}c@{}}Beam\\ {[}mas $\times$ mas, deg{]}\end{tabular}} & Array \\ \hline
1.6                                                                                      & 2005-03-09                         & $16.90 \times 12.40$, -87                                                                                & EVN                                                                                      \\
5.0                                                                                      & 1994-11-15                         & $2.48 \times 1.29$, -5                                                                                   & VLBA                                                                                       \\
5.0                                                                                      & 1995-10-28                         & $2.21 \times 1.34$, -3                                                                                   & VLBA                                                                                      \\
5.0                                                                                      & 1996-05-10                         & $3.65 \times 2.56$, -18                                                                       & VLBA $^{(*)}$, Y                                                                                       \\
5.0                                                                                      & 1996-10-07                         & $2.31 \times 1.35$, -5                                                                                   & VLBA, Y                                                                                       \\
15.4                                                                                     & 2019-08-23                         & $0.82 \times 0.42$, -5                                                                                                          &  VLBA                                                                                          \\
22.3                                                                                     & 2007-12-02                         & $0.94 \times 0.25$, -22                                                                         & VLBA $^{(**)}$, Y, \\ & & & GB, EF                                                                                      \\
22.3                                                                                     & 2018-11-24                         & $0.59 \times 0.27$, -11                                                                         &  VLBA, Y, EF                                                                                     \\
43.2                                                                                     & 2008-02-03                         & $0.27 \times 0.14$, -20                                                                                  & VLBA$^{(*)}$, Y, 
\\ & & &  GB, EF          \\
43.2                                                                                     & 2018-11-24                         & $0.30 \times 0.16$, -13                                                                                  & VLBA, Y, EF          \\
\hline
\end{tabular}
\label{tab:beam}

\vspace{0.3cm}

{\bf{Notes.}} Column 1: Frequency;  Column 2: Date of observation; Column 3: Beam FWHM and position angle; Column 4: Array: VLBA – Very Long Baseline Array: $^{(*)}$ no data from Mauna Kea, $^{(**)}$ no data from Saint Croix; Y – Very Large Array; EVN – European VLBI network; EF – Effelsberg; GB – Green Bank;
\end{table}

In our analysis we assume that the approaching jet is oriented at an angle $\theta=38\degree$ with respect to the line of sight, following the results of the 5 GHz Very Large Array (VLA) study presented by \citet{Canvin_2005}, and within the range of $30\degree$-$40\degree$ derived by \citet{Giovannini_2001} based on VLBI data. While noticing that a larger angle of the order of $50\degree$ has been derived in the works of \citet{Laing_2014} and \citet{Park_2021}, we will provide a further constraint in Sect.\ \ref{sec:Jet_acc} in support of our selection of $\theta=38\degree$.

With respect to B21, a more recent estimate of the black hole mass in NGC\,315 is assumed in this work. Through CO observations performed with the Atacama Large Millimeter/submillimeter Array (ALMA), the black hole mass  was estimated to be $M_\mathrm{BH} = (2.08 \pm 0.01)^{+0.32}_{-0.14} \times 10^9 M_\odot$ \citep{Boizelle_2021}. Under this assumption, and for $\theta=38\degree$, one projected milliarcsecond correponds to $2701\,R_{\rm S}$ de-projected.  


The opacity shifts necessary to properly align maps at different frequencies are assumed from B21 (Table 4), who performed 2D cross-correlation analyzes between different pairs of frequencies. All distances are expressed with respect to the 43 GHz core, assumed to coincide with the actual position of the black hole. 
This assumption is supported by previous observational results, showing that the mm-core in radio galaxies lies very close ($<100 R_{\rm S}$) to the true position of the central black hole \citep[e.g.,][]{Hada_2011, Baczko_2022}. 

We analyze the CLEAN maps through a pixel-based analysis using a Python code. This requires, as input, maps aligned along the y axis restored with a circular beam,
and divides both the jet and the counter-jet in one-pixel width slices oriented perpendicularly to the jet direction.
The intensity profiles in the slices are subsequently fitted with a single 1D Gaussian profile by means of the Levenberg-Marquardt algorithm ({\tt LevMarLSQFitter} in {\tt Astropy}).
Pixels with value lower than $3 \sigma_{\mathrm{rms}}$ are discarded, to avoid the contamination from the background noise at the edges of the jet.
The code stops its iteration when the most luminous pixel in the slice has a brightness lower than $10  \sigma_{\mathrm{rms}}$.
Finally, the fitted Gaussian profiles are integrated to obtain the flux density in each slice.
The integrated fluxes, together with the full width at half maximum (FWHM) of the Gaussian profiles, are stored and used to compute the jet to counter-jet ratio (Sect.\ \ref{sec:Jet_acc}) and the jet opening angle as a function of distance from the core (Sect.\ \ref{sec:Jet_shape}). 
We point out that in our analysis, we discard the pixels within a beam size region centered in the image core due to the inability to distinguish between nuclear and jet emission.

\section{Results} \label{sec:Results}

\begin{figure*}[h]
\begin{multicols}{2}
    \includegraphics[width=\linewidth]{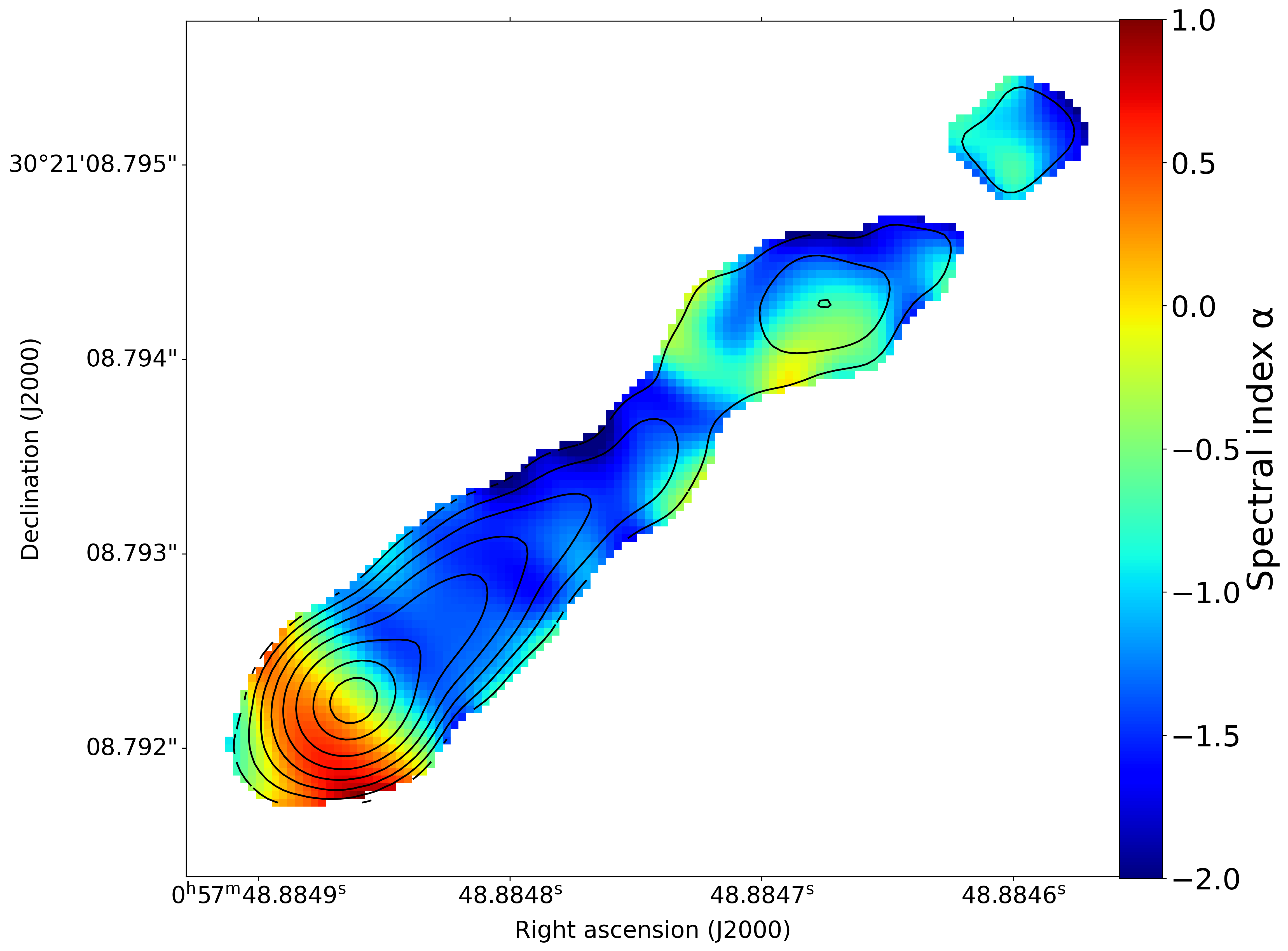}\par
    \includegraphics[width=\linewidth]{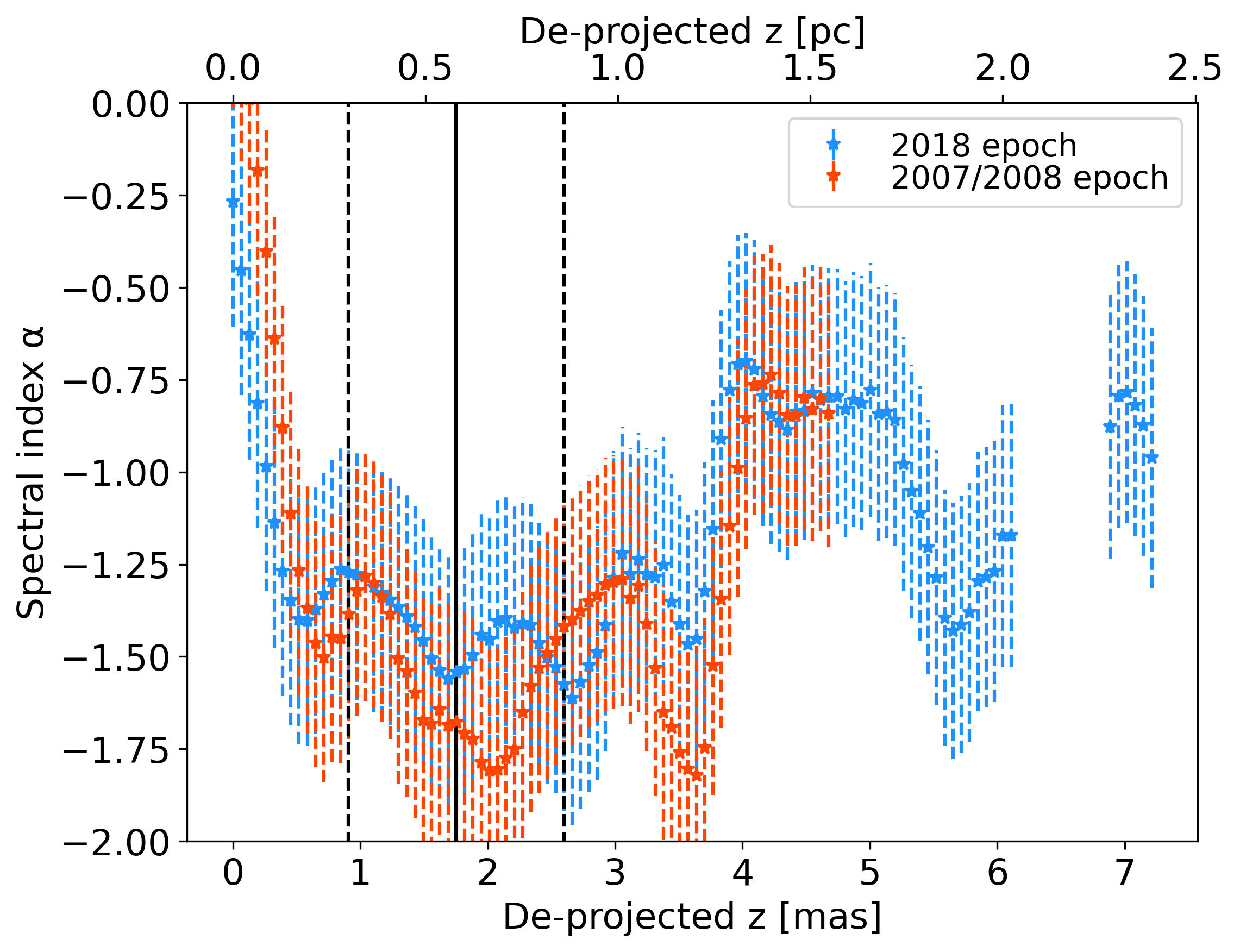}\par
\end{multicols}
\caption{Left panel: spectral index map between the frequencies 22 GHz and 43 GHz. The contours outline the 43 GHz continuum emission. Right panel: average spectral index along one-pixel slices as a function of distance from the core for two different epochs.
The two profiles are in agreement within the errors.
The black vertical lines demarcate the jet shape transition distance with its uncertainty. Within the collimation region, the spectral index rapidly decreases down to values of $\alpha\sim-1.5$, while it rises to typical values of $\alpha\sim-0.75$ beyond the shape transition distance.}
\label{fig:spectral_map}
\end{figure*}

\subsection{Spectral index} \label{sec:Spectral}

We use our 22 GHz and 43 GHz images to investigate the spectral index ($S_\nu \propto \nu^{\alpha}$) from sub-parsec to parsec scale.
We consider our simultaneous observations from 2018, as well as the 2007-2008 observations (see Table \ref{tab:beam}), which are separated by 63 days. In the latter case the analysis is aimed at a consistency check for the 2018 data set, taking into account that in a low-power radio galaxy like NGC\,315 we do not expect strong variability over such a relatively short period. 
Indeed, in radio galaxies the flux variability is softened by a reduced relativistic beaming due to the large viewing angles. 
Moreover, even small flux density variations statistically occurs on time scales of 1-2 years \citep{Hovatta_2007}.
The paired maps were produced by limiting the uv-range to the same interval, and were convolved with the same equivalent circular beam. Moreover, the core shift between the two frequencies was applied as explained in Sect.\ \ref{sec:Data_set}.
The uncertainties are computed by propagating the errors on the flux calibration, set to $5\%$ for the 22 GHz map and to $10\%$ for the 43 GHz map, together with the rms noise (for the image noise values, we refer to B21, Table 1). 
Pixels with value lower than $10 \sigma_{\mathrm{rms}}$ were masked.

In the case of the 2018 data set, the analysis was performed twice, first by considering the maps obtaining using the full HSA array and the second after reproducing the images using only the VLBA. This second approach was aimed at minimizing the uncertainties possibly introduced in the amplitude calibration by the two non-VLBA antennas (Effelsberg and the phased-VLA). We obtained comparable results in the two cases, with slightly steeper values determined when the full array was considered. 
In Fig.\ \ref{fig:spectral_map}, left panel, we show the spectral map from the 2018 VLBA data.
The contours describe the 43 GHz continuum emission.
The core structure is marked by the red patch which identifies the optically thick region, with inverted spectral index values up to $\alpha \sim 0.7$.
The average spectral index evolution along one-pixel slices is presented in Fig.\ \ref{fig:spectral_map} right panel for both epochs, showing 
a remarkable agreement.
In the optically thin region the spectral index drops down to steep values of $\sim -1.5$ using the VLBA-only data. Slightly steeper values down to $\alpha \sim -1.75$ are observed using the full-HSA data as well as based on the 2007-2008 epochs.
The steep index region extends up to $\sim 1 \ \mathrm{pc}$, a distance comparable to the extent of the jet parabolic region proposed by B21.
Beyond this region, the spectrum become flatter and reaches typical values of $\alpha \sim -0.75$.
Synchrotron losses reflecting the presence of strong magnetic fields may explain the steep spectrum we observe upstream of the shape transition distance. We note that, in general, the spectral properties in the jet acceleration and collimation region are unknown, as they have not yet been investigated in observational studies. A detailed analysis considering a broader multi-frequency data set will be necessary in the case of NGC\,315, and of other sources that can be examined on similar scales. 


\subsection{Jet speed profile} \label{sec:Jet_acc}
We constrain the intrinsic jet speed by considering the jet to counter-jet intensity ratio $R$.
Under the assumption of intrinsic symmetry the jet and counter-jet, the observed flux density differences are only due to the Doppler boosting of the incoming jet and de-boosting of the receding one.
We compute the flux density ratio between slices which are equidistant with respect to the 43 GHz core position. The intrinsic jet speed $\beta$ may be expressed as:
\begin{equation}
    \beta = \frac{1}{\mathrm{cos}(\theta)} \frac{R^{1/p} - 1}{R^{1/p} + 1}
    \label{eq:J_CJ}
\end{equation}
where $p = (2 - \alpha)$ and $\alpha$ is the spectral index \citep{Urry_1995}.
To compute Eq.\ \ref{eq:J_CJ} we assume, in the case of the 22 GHz and 43 GHz data, the spectral index values obtained in Sect. 3.1 for the 
VLBA-only maps,
while at the lower frequencies we adopt average values following the results of \citet{Park_2021}.
Namely, we set $\alpha = -0.80 \pm 0.20$ for the 5 and 15 GHz data and $\alpha = -0.60 \pm 0.30$ for the 1 GHz data. 
The assumption of an average value for the index does not affect our conclusions due to the small oscillations of the spectral index around the values assumed.

\begin{figure}[t]
    \centering
    \includegraphics[width=\linewidth]{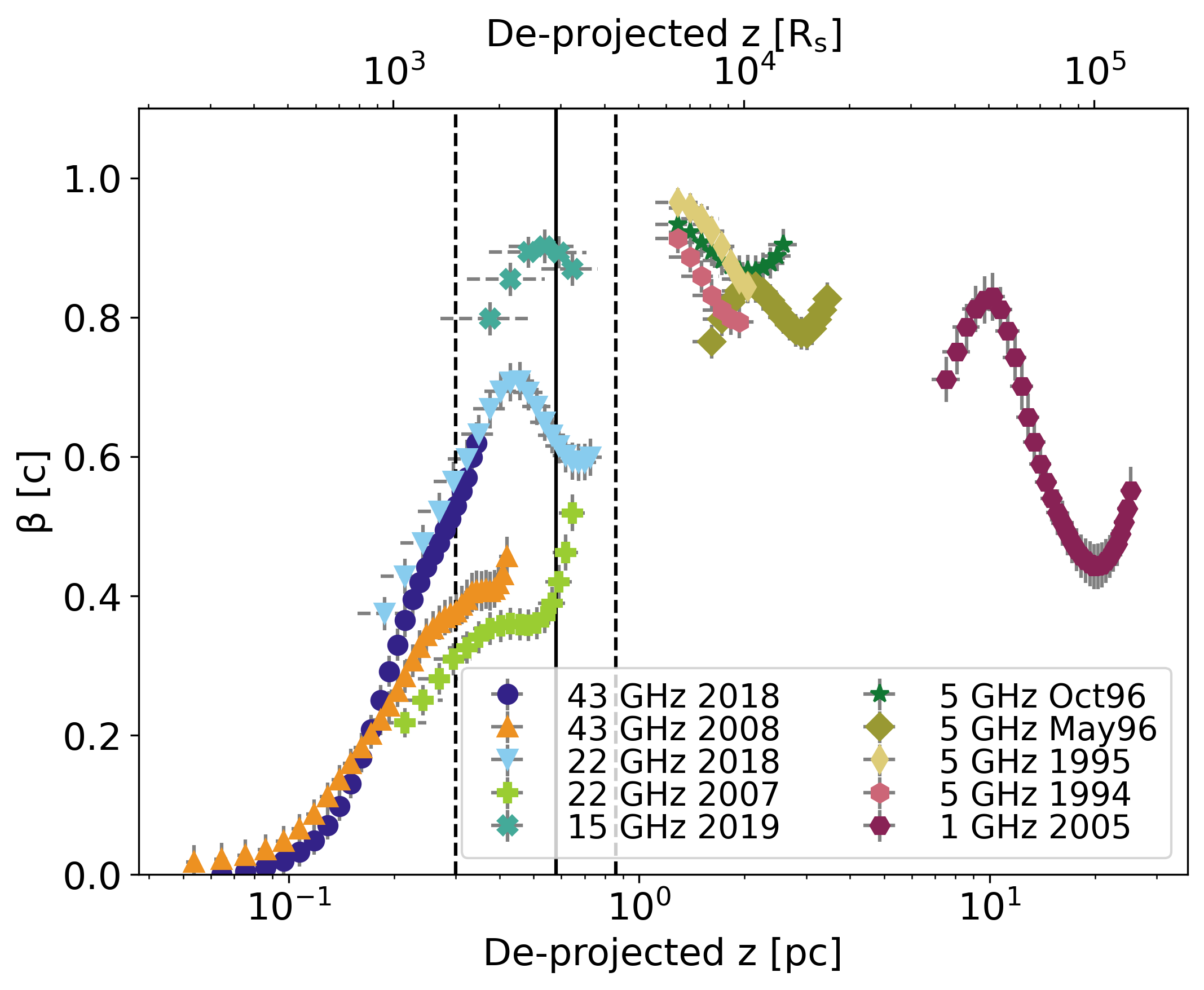}
    \caption{Jet intrinsic speed profile based on the analysis of the jet to counter-jet intensity ratio. The solid black vertical line indicates the transition distance of $\mathsf{z} = 0.58 \pm 0.28 \, \mathrm{pc}$ (accordingly to B21) with the uncertainties enclosed by the dashed black lines. The bulk of the jet accelerates on sub-parsec scales, concurrently with the parabolic jet expansion. The highest velocity of $\beta = 0.95 \pm 0.03$ is reached around 1 pc.}
    \label{fig:J_CJ_ratio}
\end{figure}

The uncertainties on $\beta$ are computed by propagating the error on the spectral index and the error on the intensity ratio. 
The uncertainties on the intensity are computed taking into account both systematic errors and uncertainties associated with the method. 
The systematic errors, including errors on the calibration and imaging of the data, are assumed to introduce a fixed error on the final intrinsic ratio equal to $10\%$ of its value.
The uncertainties associated to the slicing method are predominantly related to the position angle of each slice.
To quantify them, we vary the position angle in the range $\pm 3\degree$, with an interval of $1\degree$, compute the standard deviation in each slice for each angle and sum them in quadrature.
\citet{Pushkarev_2017} used a similar method spanning a range of $\pm 15\degree$, but since the direction of NGC\,315 is well-defined (as seen from the ridgeline in Fig.\ \ref{fig:ridgeline}) we span an interval of 6\degree \, in total, as with a shift of 3\degree \ the jet is clearly misaligned.
The $\beta$ profile for each analyzed map is shown in Fig.\ \ref{fig:J_CJ_ratio}.
The lack of data points around 1 and 4 pc results from the inner pixels discarded in the 1 GHz and 5 GHz maps, following the procedure explained in Sect.\ \ref{sec:Data_set}.

We observe a very fast accelerating jet on sub-parsec scales, where terminal intrinsic velocities of $\beta \sim 0.9$ are reached already around $\sim 0.6 \ \mathrm{pc}$.
Such fast acceleration is inferred in all the maps that sample the parabolic region and is discussed in Sect.\ \ref{sec:jet_acc_grad}. The acceleration observed in the 22 GHz - 43 GHz maps from 2018 is steeper than the one observed at the same frequencies in 2007-2008. 
This may indicate that the gradient is time-variable, due to variable physical conditions at the jet base.
Moreover, we notice how the 22 GHz - 2007 map is slightly misaligned with respect to the 43 GHz - 2008 map. 
This is not surprising, since
the parsec-scale core location can vary with time, even drastically during flares, due to the varying opacity at the jet base \citep{Plavin_2019}.

On the scales sampled by the multi-epoch observations at 5 GHz we observe no acceleration, but a small deceleration between one - two parsec. 
The velocity inferred from four different epochs at 5 GHz varies between $\beta \sim 0.8-0.95$.
The 1 GHz data indicate a deceleration on scales of $\sim 10 \, \mathrm{pc}$, down to $\beta \sim 0.4$, and a similar behavior has been reported for the 1 GHz data points by \citet{Park_2021}.
However, it is unclear whether we are observing a true deceleration of the bulk flow. 
Indeed, intrinsic velocities of $\sim 0.9 \mathrm{c}$ are measured on kiloparsec scales as well, suggesting that the terminal speed reached by the jet on sub-parsec scales is maintained up to large distances,
with the deceleration actually found to occur at distances between 8 and 18 kpc, where the speed decreases to $\beta \sim 0.4$  \citep{Canvin_2005, Laing_2006}.
One possibility for reconciling these results is that the VLBI jet is not uniform but stratified.
The 1 GHz emission may be produced in a wider sheath at the jet edges, where the bulk speed might be lower, while the higher frequency emission may be produced by more compact regions closer to the jet axis. This interpretation may be supported by the observation of a frequency-dependent opening angle, which we discuss in the next Section. 

The jet break distance, that is where the transition from parabolic to conical profile occurs, of $\mathsf{z}_\mathrm{br} = 0.58 \pm 0.28 \, \mathrm{pc}$ (B21), is emphasised by the black vertical lines in Fig.\ \ref{fig:J_CJ_ratio}.
From a visual inspection of the speed profile, the acceleration takes place within the jet parabolic region, while approximately constant speeds are observed at the start of the conical jet expansion.  
This is expected when the acceleration is mainly driven by the conversion of Poynting flux into kinetic energy of the bulk flow \citep{Komissarov_2012}.

Using Eq.\ \ref{eq:J_CJ} with $\beta = 1$ and the maximum ratios $R$ observed, we can set an upper limit to the viewing angle.
We use the highest ratios detected at 5 GHz and 15 GHz, since they are the maps from which we infer the highest velocities.
The highest ratios at 5 GHz, starting from the newest map, are $R_1 = 194 \pm 22$, $R_2 = 89 \pm 11$, $R_3 = 267 \pm 32$, $R_4 = 161 \pm 20$, while at 15 GHz is $R_5 = 145 \pm 22$.
From them, we find an average maximum viewing angle of $\theta_\mathrm{max} = 44 \pm 4 \degree$, in agreement with our assumption of $\theta = 38\degree$.

\subsection{Jet opening angle} \label{sec:Jet_shape}

In this section we investigate the intrinsic jet opening angle profile using the observations reported in Fig.\ \ref{fig:J_CJ_ratio}, as well as the  stacked images at 5 GHz and 15 GHz.

\begin{figure}[t]
  \includegraphics[width=\linewidth]{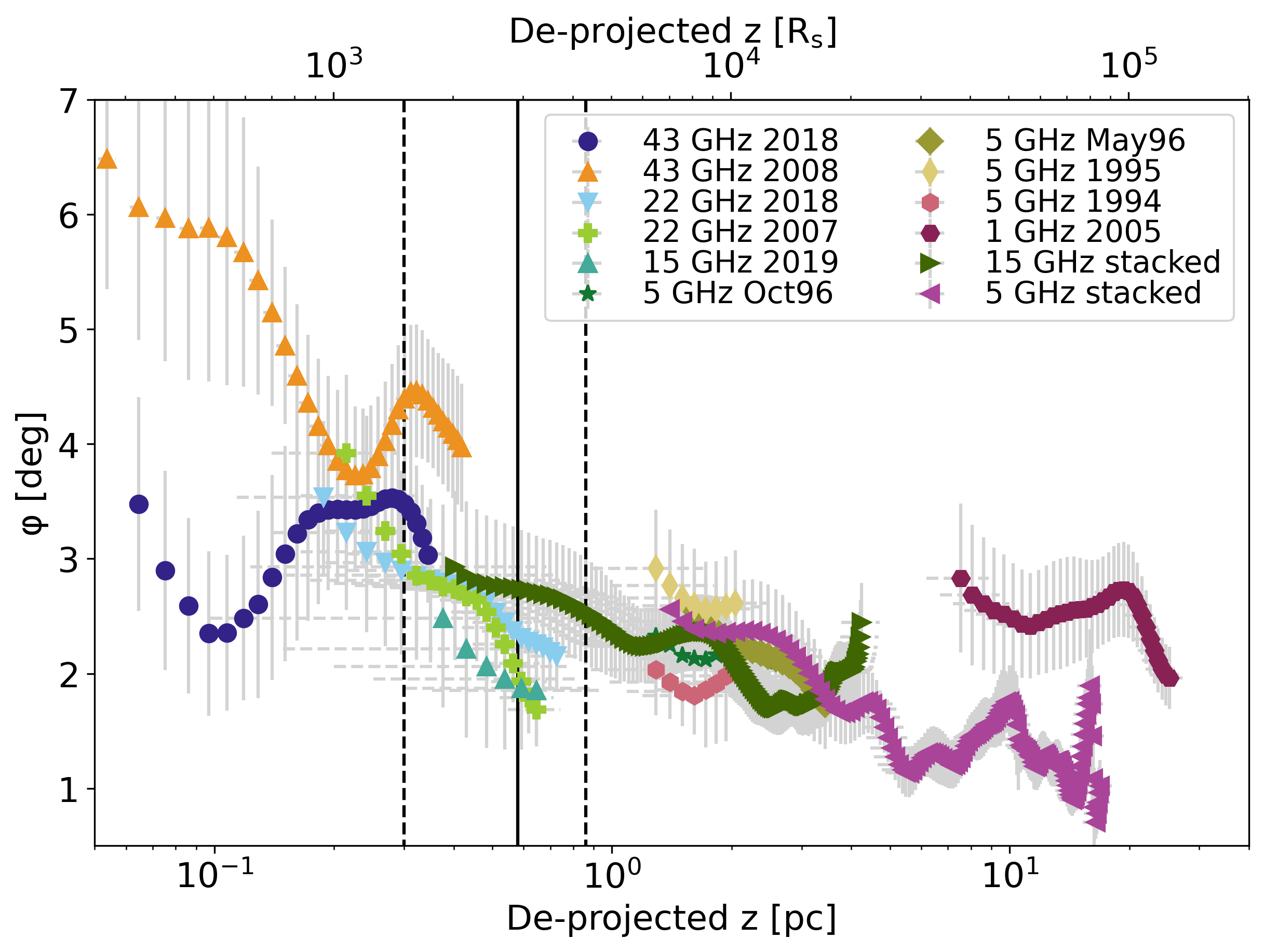}
  \caption{Intrinsic half opening angle in the frequency range 1 GHz - 43 GHz. The angle varies between $\sim 1\degree$ and $\sim 6\degree$. A  decrease is observed within one parsec, while an approximately  constant value of $2\degree-3\degree$ is determined downstream. Note the discrepancy between the results at 1 GHz and 5 GHz, which may indicate that the jet cross-section is not fully detected at 5 GHz beyond 4 pc.}
  \label{fig:opening_angle}
\end{figure}

To study the opening angle $\phi_\mathrm{app}$ we need to first obtain the de-convolved transverse jet width, computed as $w = \big(w_m^2 - \theta_b^2\big)^{1/2}$, where $w_m$ is the width of the fitted Gaussian profile and $\theta_b$ is the full width at half maximum of the equivalent circular beam, defined as $\theta_b = \sqrt{\theta_{min} \cdot \theta_{max}}$ with $\theta_{min}$ and $\theta_{max}$ being the minor and major beam axis, respectively.
The uncertainties on the width are computed as described in Sect.\ \ref{sec:Jet_acc} for the intensity ratio.

Finally, from our knowledge of both the jet width and the jet viewing angle, we compute the apparent jet opening angle as
\begin{equation}
    \phi_{\rm app} = 2 \, \mathrm{arctan} (0.5w/\mathsf{z}) \ ,
\end{equation}
where $\mathsf{z}$ is the projected distance from the core along the jet, as well as the intrinsic half opening angle $\phi$ as
\begin{equation}
    \phi = \mathrm{arctan} \big[\mathrm{tan} \, (\phi_{\mathrm{app}}/2) \, \mathrm{sin} \, \theta \big] \ .
\end{equation}
The intrinsic half opening angle profiles are illustrated in Fig.\ \ref{fig:opening_angle}.
For the single epoch maps, we plotted the points up to the same jet extension as in Fig.\ \ref{fig:J_CJ_ratio}, while for the stacked images we show the full jet.
The uncertainties are computed by propagating the errors obtained on the intrinsic jet width.
The intrinsic half angle is on average $\phi \sim 2\degree - 3\degree$ and spans a range between $\sim 1\degree$ and $\sim 6\degree$, in agreement with the typical values for radio galaxies \citep{Pushkarev_2017}. While results for most frequencies and distances agree quite well with each other, we notice large discrepancies on sub-parsec scales between the two 43 GHz epochs and at a distance larger than $\sim5$ de-projected pc, where the 5 GHz jet appears narrower than at 1 GHz. 
The sub-parsec scales discrepancy may arise due to an intrinsic jet variability across the epochs, combined with the inability to recover the full jet cross-section in the 43 GHz 2018 epoch, where the inner part of the jet is indeed less bright than the 2008 epoch (see Fig.\ \ref{fig:43GHz}).
The narrowed jet at 5 GHz may indicate that at this frequency, even in the stacked image, part of the jet at its edges is too faint to be imaged, so we are not able to recover the full jet cross-section. This of course would also have an impact on the determination of the global jet shape, which was found to be not perfectly conical (power-law index of $\sim 0.84 \pm 0.04$) on those scales in B21. These results suggest that the jet may, in fact, be perfectly conical, but part of its emission may be lost in the noise at 5 GHz. Moreover, a frequency-dependent opening angle may be consistent with the discrepancy in the speeds discussed in Sect. 3.2, and explain the apparent deceleration observed at 1 GHz. 

\subsection{$\Gamma \phi$ profile} \label{sec:Gamma-Theta}

The product $\Gamma \phi$, where $\Gamma = 1/\sqrt{1 - \beta^2}$ is the bulk Lorentz factor, is of a great interest since it carries information on the causal connection between the edges and the symmetry axis of the flow, with $\Gamma \phi \lesssim 1$ meaning that the jet is causally connected \citep{Clausen-Brown_2013}.

In magnetically-accelerating relativistic outflows, the bulk acceleration requires the jets to be causally connected, with $\Gamma \phi \sim 1$, while a product higher than one can be observed if the dissipation of thermal energy contributes to the jet acceleration \citep[e.g.,][]{Komissarov_2009}.
Moreover, the product is connected to physical aspects of the accretion disk.
In particular, $\Gamma \phi$ is related to the dimensionless magnetic flux $\phi_\mathrm{BH}$, which is an important indicator of the magnetic activity in the core.
When $\phi_\mathrm{BH}$ is of the order of few units, a SANE disk is expected, while when it is of the order of tens and especially when it reaches the saturation level $\sim 50$, we expect the disk to achieve a MAD state \citep{Tchekhovskoy_2011, Narayan_2012}.
The dimensionless flux is defined as:
\begin{equation}
    \phi_\mathrm{BH} = \Phi / (\sqrt{\dot{M} c} \, r_g )
    \label{eq:dimensionless_flux}
\end{equation}
where
\begin{itemize}
    \item $\Phi$ is the total magnetic flux in the jet;
    \item $\dot{M}$ is the accretion rate;
    \item $c$ is the speed of light;
    \item $r_\mathrm{g} = GM_\mathrm{BH}/c^2$ is the gravitational radius,
\end{itemize}
and in the surroundings of the central black hole the quantity $\Gamma \phi$ is related to $\phi_\mathrm{BH}$ via $\phi_\mathrm{BH} = (52 \pm 5) \Gamma \phi$ \citep{Zamaninasab_2014}.

\begin{figure}[t]
  \resizebox{\hsize}{!}{\includegraphics{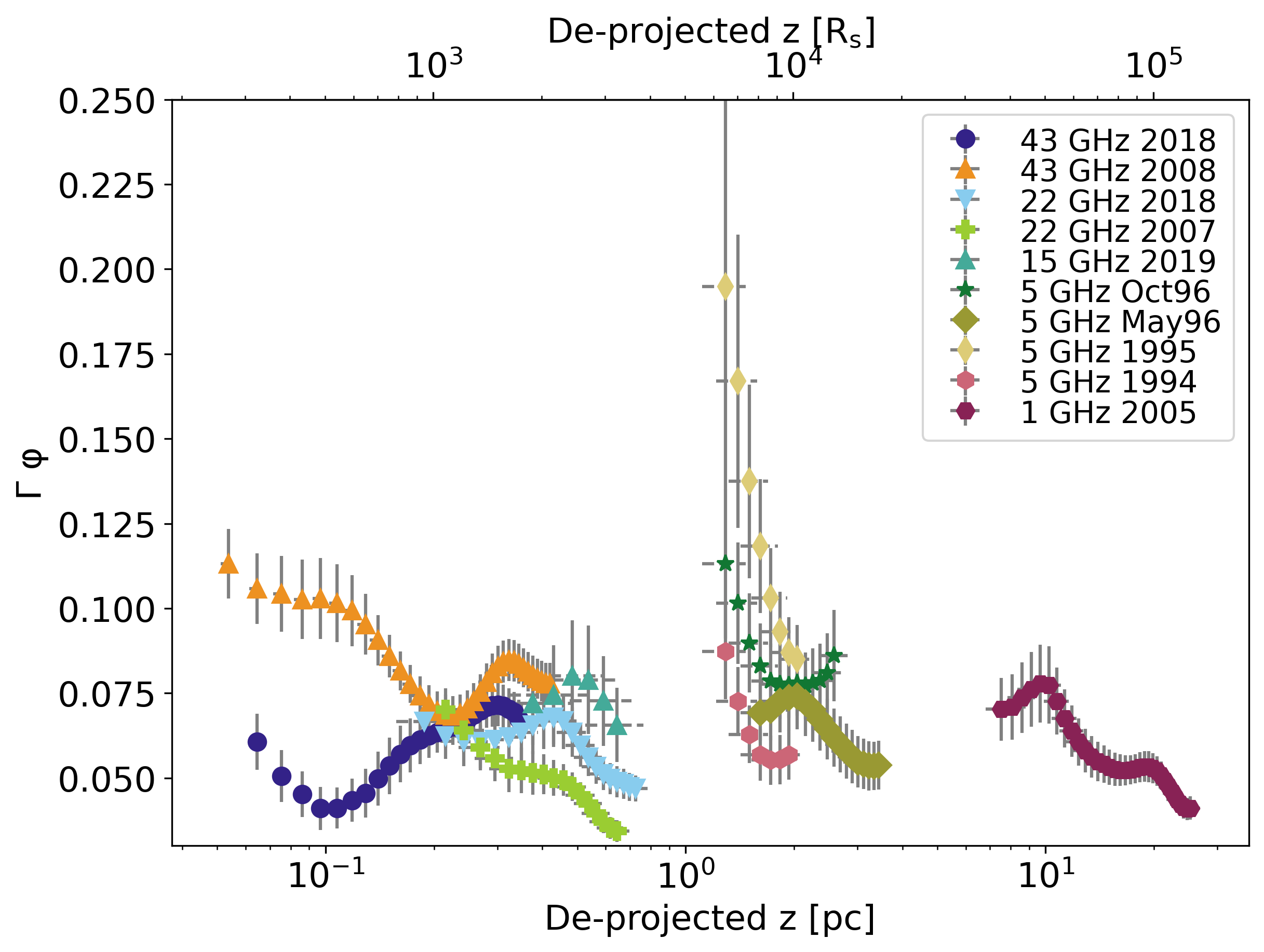}}
  \caption{$\Gamma \phi$ product as a function of distance. The profile is fairly constant, with hints of a decreasing trend  downstream the jet on scales smaller than 0.1 pc and 0.2 pc seen from the 43 GHz 2018 and 43 GHz 2008 data points, respectively. The average product is $\Gamma \phi{\sim} 0.07$.
  }
  \label{fig:G_T}
\end{figure}

Here, we compute the $\Gamma \phi$ profile using our information on the jet speed (Sect.\ \ref{sec:Jet_acc}) and on the half opening angle (Sect.\ \ref{sec:Jet_shape}).
The uncertainties are computed by propagating the respective errors.
The resulting profile is shown in Fig.\ \ref{fig:G_T}.
We find a fairly constant product, varying between ${\sim}$0.04 and ${\sim}$0.2 with an 
average value of $\sim 0.07$, consistent with the median products found in large samples, $\Gamma \phi = 0.13-0.17$ \citep{Jorstad_2005, Pushkarev_2009, Pushkarev_2017}.
Similar values are observed in M87, specifically $\Gamma \phi \sim 0.046$ at 0.3 pc and $\Gamma \phi \sim 0.044$ at 4 pc \citet{Nohkrina_2019}.
Hints of a decreasing trend downstream the jet are given on scales smaller than 0.1 pc and 0.2 pc by the 43 GHz 2018 and 43 GHz 2008 data points, respectively.
These observational findings deviate from the theoretical expectation that $\Gamma \phi \sim 1$ in a purely magnetically-driven accelerated jet \cite[e.g.,][]{Komissarov_2009}.
In the discussion (Sect.\ \ref{sec:Discussion}) we indirectly infer a value of $\Gamma \phi$ in the disk region by means of our estimate of the dimensionless magnetic flux.
The implications on the evolution of the product are discussed in Sect.\ \ref{sec:G_T_disc}.

\section{Discussion} \label{sec:Discussion}

In the following, we apply theoretical models to our observational data to explore the nature of the accretion disk and of the outflow in NGC\,315.
In particular, we aim at modeling  the jet acceleration seen in Fig.\ \ref{fig:J_CJ_ratio}, the magnetic field down to the jet base, and at investigating whether the properties of the VLBI jet are consistent with the expectations for a jet powered by the rotational energy of a black hole fed by an accretion flow in MAD state. Such a test will be performed starting from two independent observables: the location of the jet shape transition and the mass accretion rate. To define the main properties of the accretion flow, we will make use of optical and X-ray observational constraints.

\subsection{Spin and magnetic flux inferred from the shape transition distance} \label{sec:BH_spin}

First, we use the information on the jet shape transition distance to estimate two fundamental parameters of the nuclear region, the black hole spin and the total magnetic flux threading the accretion disk.
The former is used to evaluate the jet power (see Sect.\ \ref{sec:jet_power}), while the latter is compared with the expected flux in case of a MAD.
To do that, we apply the model proposed by \citet{Nohkrina_2019, Nokhrina_2020}.
The model assumes equilibrium between the ambient medium and the jet edges, with the former being balanced only by the jet thermal pressure at the boundary.
An electric current is assumed to be mainly locked within the bulk jet volume and only a residual fraction of it is left in the outer jet boundary.
Following \citet{Beskin_2017}, for a non-relativistic boundary flow the conservation of the total pressure allows us to connect the pressure balance at the jet boundary with the electric current inside the jet. 
The boundary pressure is expressed as function of the jet width.
\citet{Kovalev_2020} showed that when the jet energy regime switches from magnetically dominated to equipartition, that is when the Poynting flux is equal to the kinetic energy of the bulk, this coincides with a transition between two power laws in the jet boundary pressure profile due to different behaviors of the electric current in the two regimes.
Therefore, when the acceleration of the bulk flow saturates the residual electric current, which diminished due to the transformation of the Poynting flux, the current impacts the pressure at the jet boundary, leading to the occurrence of a break in the jet shape. 
By knowing the jet properties at the transition, this model enables us to obtain information on the central engine, given the observed transition distance and assuming that the external ambient pressure evolves with a single power law profile $P(\mathsf{z}) \propto \mathsf{z}^{-b}$.
The spin is expressed via the dimensionless parameter $a_* = J/M_\mathrm{BH} \, \in \, [0,1]$, where $J$ is the angular momentum of the central black hole, as:
\begin{equation}
    a_* = \frac{8 \, (r_\mathrm{g}/R_\mathrm{L})}{1 + 16 \, (r_\mathrm{g}/R_\mathrm{L})^2}
    \label{eq:spin}
\end{equation}
where the light cylinder radius $R_\mathrm{L}$: 
\begin{equation}
    R_\mathrm{L} = \frac{r_{\mathrm{br}}}{d_*(\sigma_\mathrm{M})}.
\end{equation}
Here, the light cylinder radius is the intrinsic length scale in MHD models and $d_*$ is the non-dimensional jet radius that depends on the initial magnetization of the jet $\sigma_\mathrm{M}$, defined as the ratio between the Poynting flux to the plasma rest mass energy flux.
\footnote{The values for $d_*$ are taken from \citet{Nohkrina_2019}, Table 1.}
In a cold outflow the jet accelerates efficiently up to the equipartition regime, where approximately half of the initial electromagnetic energy is converted into kinetic energy of the
particles
\citep[see][and references therein]{Nohkrina_2019}. 
Thus, $\sigma_\mathrm{M}$ is roughly twice the maximum observed Lorentz factor. 
From our analysis on the jet kinematics (see Sect.\ \ref{sec:Jet_acc}), we constrain a maximum Lorentz factor of $\Gamma \sim 3-4$.
Therefore, a initial magnetization of the order of ${\sim}10$ seems reasonable in our case.
However, the $\Gamma$ observed in our study is unlikely to be the true maximum reached by the jet. 
Indeed, if the jet has a spine-sheath structure, the large viewing angle of NGC\,315 implies that the radio emission is mainly associated with the slower sheath, thus we may be unable to directly probe the central spine. 
Since NGC\,315 is a low-excitation FRI radio galaxy, a realistic assumption for the speed of the spine can rely on VLBI studies of the blazar parent population, formed by BLLac objects. 
These typically show $\Gamma_\mathrm{max} \sim 10$
\citep[e.g.,][]{Hovatta_2009} which leads us to consider $\sigma_\mathrm{M} \sim 20$ as a reasonable upper limit for our analysis.
For the jet radius at the break we assume $r_{\mathrm{br}} = 0.036 \pm 0.013 \, \mathrm{pc}$ following the results in B21.
Then, for $\sigma_\mathrm{M} = 10$ and  $\sigma_\mathrm{M} = 20$, Eq.\ \ref{eq:spin} yield to black hole spins $a_* = 0.86^{+0.13}_{-0.14}$ and $a_* = 0.99^{-0.04}_{-0.08}$, respectively. 
The asymmetric errors arise due to the particular relation of the spin $a_*$ with respect to the radius at the break $r_\mathrm{br}$, with the former first increasing and then decreasing with increasing radius.
This model points toward a fast rotating black hole, which is necessary to efficiently launch powerful jets from a MAD accretion flow 
\citep[e.g.,][]{Narayan_2021}.

The total magnetic flux in the jet is computed as \citep{Nohkrina_2019}:
\begin{equation}
    \Phi = 2 \pi R_\mathrm{L}^2 \sigma_\mathrm{M} \bigg[\bigg(\frac{\mathsf{z}_\mathrm{br}}{r_\mathrm{0}}\bigg)^{-b} \frac{P_0}{p_*(\sigma_\mathrm{M})}\bigg]^{1/2} \ \mathrm{G} \ \mathrm{cm^{2}} \, 
    \label{flux_Nokhrina}
\end{equation}
where:
\begin{itemize}
    \item $P_0$ is the external pressure in the nuclear region within a sphere of radius $r_0$;
    \item $p_*(\sigma_\mathrm{M})$ is the non-dimensional ambient pressure.
\end{itemize}
The break distance of $\mathsf{z}_\mathrm{br} = 0.58 \pm 0.28 \, \mathrm{pc}$ is inferred by B21, while the external pressure of $P_0 = 4.5 \times 10^{-10} \, \mathrm{dyn} \, \mathrm{cm^{-2}}$ at 1 arcsec from the core was determined by \citet{Worrall_2007} based on X-ray Chandra data. We note that the radius within which the pressure measurement was extracted ($\sim 0.3$\,$\rm kpc$) is larger than the estimated Bondi radius 
\footnote{The Bondi radius is estimated following \citet{Russell_2015}.}
($\sim 0.15 \, \mathrm{kpc}$ for a black hole mass of $M_\mathrm{BH} = 2.08 \times 10^9 M_\odot$), but not by a large factor.
The value of b may vary between 1.0 and 2.0 \citep[see discussion in Sect.\ 3,][]{Nohkrina_2019}. 
In the first case we find $\Phi = (4.6 \pm 3.5) \times 10^{31} \, \mathrm{G} \, \mathrm{cm^2}$, while in the second $\Phi = (1.1 \pm 0.9) \times 10^{33} \, \mathrm{G} \, \mathrm{cm^2}$.
These two values set the upper and lower limit on the total magnetic flux accordingly to the variation of $b$.
The magnetic flux depends weakly on the initial magnetization $\sigma_\mathrm{M}$.

We now compare the jet magnetic flux obtained via Eq.\ \ref{flux_Nokhrina} with the expected one in the MAD scenario.
The latter is computed as \citep{Zamaninasab_2014}:
\begin{equation}
    \Phi^\mathrm{exp} = 2.4 \times 10^{34} \, \bigg[\frac{M_\mathrm{BH}}{10^9 M_\odot}\bigg] \, \bigg[\frac{\dot{M} c^2}{3.15 \times 10^{47} \mathrm{erg \, s^{-1}}}\bigg]^{1/2} \ \mathrm{G} \ \mathrm{cm^{2}} \, .
    \label{eq:exp_flux}
\end{equation}
Following \citet{Heckman_2004}, we derive the mass accretion rate from the luminosity of the [O III] $\lambda 5007$ emission line as $L_\mathrm{bol} = 3500 \, L_\mathrm{O III}$, where $L_\mathrm{bol}$ is the bolometric luminosity and the bolometric correction has an uncertainty of 0.38 dex.
\citet{Ho_1993} derived a flux for the [O III] $\lambda 5007$ line of $1.17 \times 10^{-14} \mathrm{erg} \, \mathrm{s}^{-1} \, \mathrm{cm}^{-2}$, leading to $L_\mathrm{bol} = (2.44 \pm 0.93) \times 10^{43} \mathrm{erg} \, \mathrm{s}^{-1}$.
The result is in a good agreement with the bolometric luminosity of $1.9 \times 10^{43} \, \mathrm{ erg \, s^{-1}}$, obtained by \citet{Gu_2007} by integrating the SED of the nucleus of NGC\,315. 
The mass accretion rate $\dot{M} = L_\mathrm{bol}/(\eta_d c^2) =  (2.7 \pm 1.1) \times 10^{23} \, \mathrm{g/s}$ is obtained by assuming an efficiency $\eta_d = 0.1$ 
\citep[e.g.,][]{Yu_2002}
\footnote{Assuming $\eta_d = 0.4$ \citep[e.g.,][]{Zamaninasab_2014} slightly changes the quantities computed in the discussion but leaves the main conclusions unaltered.}.
Throughout the paper, the asymmetric errors are treated conservatively by computing the maximum and minimum values of the quantity and by considering them, respectively, as the upper and lower limits of the uncertainty interval.

Eq.\ \ref{eq:exp_flux} yields $\Phi^\mathrm{exp} = (1.4 \pm 0.3) \times 10^{33} \, \mathrm{G} \ \mathrm{cm^{2}}$.
The expected magnetic flux is consistent with $\Phi$ when $b \sim 2$ in the external pressure profile ($P(\mathsf{z}) \propto \mathsf{z}^{-b}$). 
Assuming $\eta_d = 0.4$, Eq.\ \ref{eq:exp_flux} would yield to $\Phi^\mathrm{exp} \sim 6 \times 10^{32} \, \mathrm{G} \ \mathrm{cm^{2}}$, consistent with $\Phi$ when $b \sim 1.8$.
For a highly magnetized accelerating jet confined by an ambient medium, a quadratic decrease of the external pressure matches indeed the observation of a jet with parabolic shape, since the jet radius is predicted to grow as $r \propto \mathsf{z}^{b/4}$ \cite[e.g.,][]{Komissarov_2009}. As in NGC\,315 we observe $r \propto \mathsf{z}^{0.45 \pm 0.15}$ in the acceleration region (B21), the index of the external pressure is expected to be $b = 1.8 \pm 0.6$.
The excellent agreement between $\Phi$ and $\Phi^\mathrm{exp}$ is a first hint of a jet launched by an established magnetically arrested disk in the nuclear region of NGC\,315.
\citet{Park_2021} found a $\sim 10$ times more extended parabolic region in NGC\,315. 
By using their jet properties at the break, we infer a slower rotation of the black hole with $a_* \sim 0.20$, while $\Phi$ and $\Phi^\mathrm{exp}$ are still in agreement.
\subsection{Exploring the $\Gamma \Phi$ profile} \label{sec:G_T_disc}

In the framework of a Blandford-Znajek jet model and 
in the hypothesis of flux-freezing approximation \citep{Blandford_1977, Zamaninasab_2014}, we can derive the dimensionless magnetic flux threading the accretion disk via Eq.\ \ref{eq:dimensionless_flux} based on our determination of the jet magnetic flux in physical units.
Using $\Phi$ and $\dot{M}$ both with $\eta_d = 0.1$ and $\eta_d = 0.4$, we obtain a dimensionless magnetic flux threading the accretion disk $\phi_\mathrm{BH} = 40 \pm 33$, consistent with the saturation level of $\sim 50$ for a magnetically arrested disk. 
Unfortunately, the large error bar does not allow us to directly discriminate between SANE and MAD.

The saturation condition $\phi_\mathrm{BH} = (52 \pm 5) \Gamma \phi$ implies $\Gamma \phi = 0.76 \pm 0.64$ at the jet base.
The large uncertainties prevent us from retrieving a solid constrain on this parameter.
However, the estimated mean value suggests a $\Gamma \phi$ product likely close to one, in agreement with the prediction for the jet launching region 
\citep[see][and references therein]{Zamaninasab_2014}, but that departs from what we observe on sub-parsec, parsec scales (see Fig.\ \ref{fig:G_T}).
If the product is actually close to 1 in the nuclear region, as expected in a cold accelerating outflow (see Sect.\ \ref{sec:Gamma-Theta}), then one of two scenarios may be invoked. $\Gamma \phi$ may start close to one and they rapidly decrease in the parabolic region (hints of a decreasing trend are given by the 43 GHz maps) down to $\sim 0.1$ at $0.1 \, \mathrm{pc}$.
Alternatively, the prediction is only valid for the very central jet spine launched in the surroundings of the black hole ergosphere, while what we infer with our observations is determined by the behavior of the external sheath.

\subsection{Jet acceleration gradient} \label{sec:jet_acc_grad}

\begin{figure}[t]
    \centering
    \includegraphics[width=\linewidth]{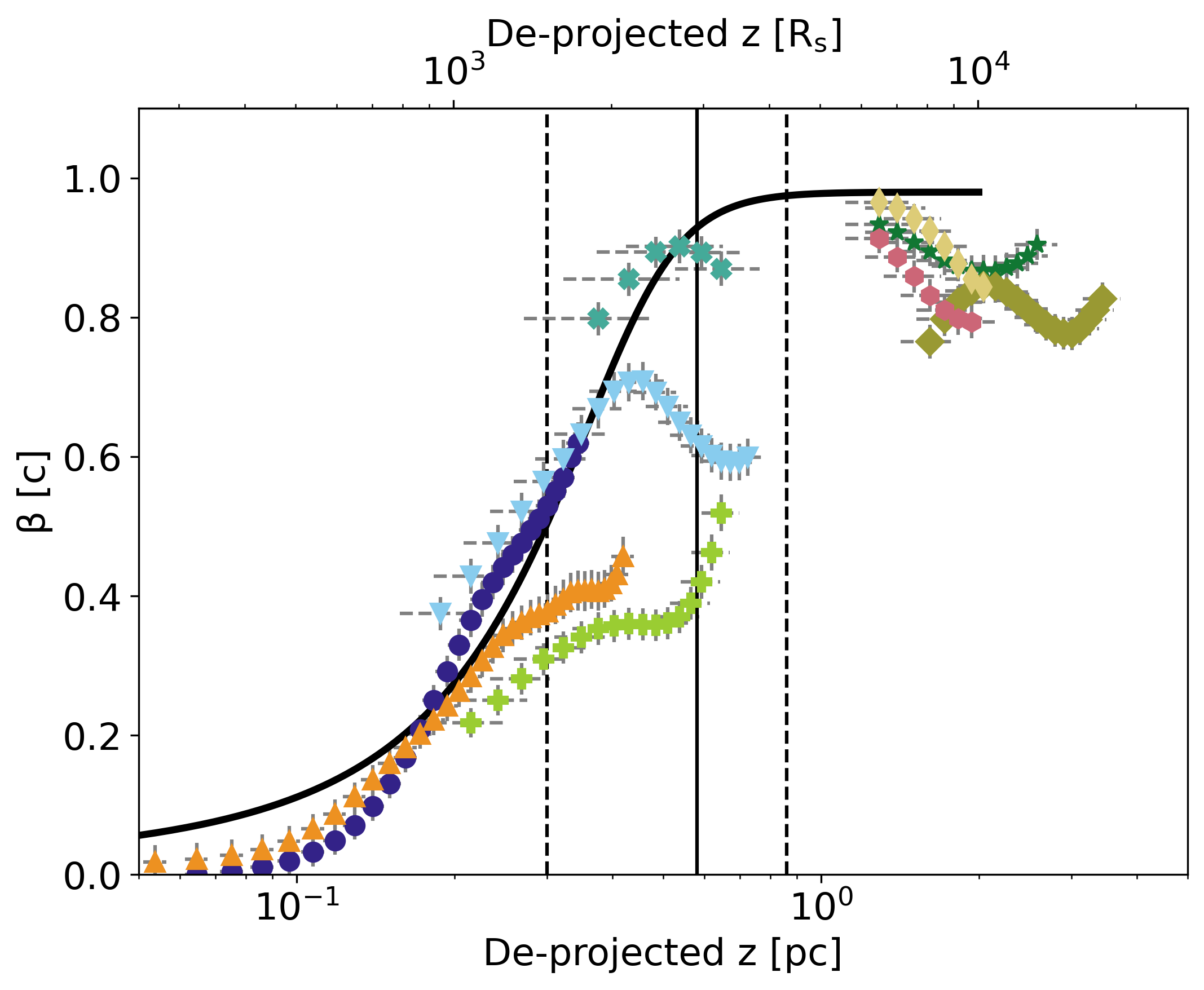}
    \caption{Zoom on the intrinsic speed profile in the distance range 0.1 - 5 pc. The black continuous line represents the modeled hyperbolic tangent function (Eq.\ \ref{eq:fit_beta}). For the label we refer to Fig.\ \ref{fig:J_CJ_ratio}.}
    \label{fig:J_CJ_fit}
\end{figure}

The jet undergoes a rapid acceleration up to $\sim 0.9 \, \mathrm{c}$ on sub-parsec scales (see Sect.\ \ref{sec:Jet_acc}).
In order to check whether the acceleration rate is consistent with the standard magneto-hydrodynamics (MHD) mechanism, we fit the speeds profile with a generic model. 
We work under the assumption of a parabolic jet shape $r = \mathrm{\lambda} \mathrm{z}^\psi$, with $\psi = 0.45$ accordingly to B21.
The scaling parameter $\lambda$ is computed from the knowledge of the jet radius at the transition distance $\mathrm{z}_\mathrm{br} = 0.58 \, \mathrm{pc}$, namely $r_\mathrm{br} = 0.036 \, \mathrm{pc}$.
To describe the observational data reported in Fig.\ \ref{fig:J_CJ_ratio}, we adopt a hyperbolic tangent function of the type $\Gamma(x) = A+B\,\mathrm{tanh}(ax-b)$, which we constrain to follow the MHD approach for a magnetically accelerating jet.
The function is used to describe the data in the Lorentz factor parameters space.
The hyperbolic tangent function mimics both the Lorentz factor growth onset on a scale of a light cylinder radius $R_\mathrm{L}$ and an acceleration saturation close to its observed maximum value $\Gamma_\mathrm{max}\sim\sigma_\mathrm{M}/2$. 
Moreover, it is linear for the argument close to zero. 
We find the constants $A$ and $B$ by setting the function limits as $1$ and $\Gamma_\mathrm{max}$. 
By ensuring that the growth rate (derivative over $r$) never exceeds $R_\mathrm{L}^{-1}$, as expected from MHD modeling  
\citep{Beskin_2006, Tchekhovskoy_2009, Lyubarsky_2009, Nakamura_2018}, we constrain the parameter $a$. 
Finally, we set $b$ so as to set $\Gamma\approx\Gamma_\mathrm{max}/2$ approximately at the distance $r\approx R_\mathrm{L}\Gamma_\mathrm{max}/2f$. 
Here, the parameter $f$ is defined as the ratio between the radius at which the maximum Lorentz factor is reached and the total jet radius. 
For models assuming a constant angular velocity \citep{Lyubarsky_2009} the maximum Lorentz factor is reached at the jet boundary ($\rho=1$), while models with a slower sheath flow provide $f\sim 0.2-0.7$ \citep{Komissarov_2007, Beskin_2017, Chatterjee_2019}. 
In our model, we use $f$ as a free parameters.
Thus we obtain
\begin{equation}
    \Gamma (\mathsf{z}) = \frac{\Gamma_\mathrm{max}+1}{2} + \frac{\Gamma_\mathrm{max} - 1}{2} \, \mathrm{tanh} \, \bigg( \frac{\mathsf{z}}{R_\mathrm{L}}\frac{2}{\Gamma_\mathrm{max}-1} - \frac{\Gamma_\mathrm{max}}{f\,(\Gamma_\mathrm{max}-1)} \bigg) \, ,
    \label{eq:fit_beta}
\end{equation}
where we associate the distance $\mathrm{z}$ with the Lorentz factor $\Gamma(f r)$.
We plot Eq.\ \ref{eq:fit_beta} using a light cylinder radius of $R_\mathrm{L} = 6.9 \times 10^{-4} \mathrm{pc}$, obtained when $\sigma_\mathrm{M} = 10$, which corresponds to $\Gamma_\mathrm{max} = 5.0$ (see Sect.\ \ref{sec:BH_spin}).
The value of $f$ that minimize the $\chi^2_\mathrm{red}$ is $f = 0.216$ ($\chi^2_\mathrm{red} = 33.3$).
The data-model comparison is done by using the 15 GHz and 22-43 GHz 2018 epochs data, since they are the ones that fully describe the acceleration region.
While the high $\chi^2_\mathrm{red}$ implies the fitting to be not statistically acceptable, here our goal is not to find a function that strictly interpolate the data with their oscillations, but instead to reconcile the general trend of the fast-acceleration with a function that follows the current MHD theories.
Fig.\ \ref{fig:J_CJ_fit} shows a zoom of the $\beta(\mathsf{z})$ plane together with the best fit.
The hyperbolic tangent function properly describes the linear acceleration on sub-parsec scales, although a discrepancy is present around $0.01 \, \mathrm{pc}$, where the model overestimates the observed velocities.
Overall, this formalism allows us to reconcile the observed fast acceleration with a magnetically-driven one in the context of the MHD theory.

Given the co-spatiality of the acceleration and collimation region (see Sect.\ \ref{sec:Jet_acc}), in our modeling  we considered the acceleration to be driven uniquely by the conversion of magnetic into kinetic energy.
We note, however, that the jets may dissipate thermal energy as well to accelerate \citep{Komissarov_2009}.
Even if our observational data suggest that the jet in NGC\,315 is magnetically dominated, the thermal acceleration may not be completely negligible and still play a role.

\subsection{Core-shift magnetic field for a conical jet and for a quasi-parabolic accelerating jet} \label{sec:B_field}

The magnetic field strength is expected to vary with distance $\mathsf{z}$ from the black hole as $B(\mathsf{z}) = B_\xi (\mathsf{z}/\xi)^\tau$, where $B_\xi$ is the magnetic field strength at $\xi$ pc \citep[see e.g.,][]{O'Sullivan_2009}.
To compute the magnetic field strength at one parsec, following the works by \citet{Lobanov_1998} and \citet{Hirotani_2005}, we define the core-position offset measure:
\begin{equation}
    \Omega_{r \nu} = 4.85 \times 10^{-9} \frac{\Delta r_{\mathrm{mas}} D_\mathrm{L}}{(1 + z)^2} \Bigg(\frac{\nu_1^{1/k_r} \nu_2^{1/k_r}}{\nu_2^{1/k_r} - \nu_1^{1/k_r}}\Bigg) \ \mathrm{pc} \ \mathrm{GHz}
    \label{eq:O_rnu}
\end{equation}
where:
\begin{itemize}
    \item $\Delta r_{\mathrm{mas}}$ is the core-shift between $\nu_2$ and $\nu_1$ (in mas);
    \item $D_L$ is the luminosity distance (in pc).
\end{itemize}
In the scenario of a conical jet with spectral index $\alpha = -0.5$ and in equipartition between the particle and magnetic field energy density, the magnetic field is obtained as:
\begin{equation}
    B_1 = 0.025 \ \bigg(\frac{\Omega_{r \nu}^3 (1 + z)^2}{\delta^2 \phi \mathrm{sin}^2\theta}\bigg)^{1/4} \ \mathrm{G} \, ,
    \label{eq:B_field}
\end{equation}
where $\delta~=~[\Gamma (1 - \beta \mathrm{cos}\theta)]^{-1}$ is the Doppler factor.
Under the same assumptions, we also expect the core-shift coefficient to be $k_r = 1$ \citep{Lobanov_1998}.
This value is slightly larger than the core-shift index assumed in this work, $k_r = 0.84 \pm 0.06$ (see Sect.\ 2.1, B21). 
The latter is indeed obtained by fitting the core-shift at all available frequencies (1-22 GHz), which sample both the parabolic and the conical jet region.
If the jet break occurs at the distance where the jet switches from a magnetically dominated to an equipartition regime \citep{Nokhrina_2020}, we should expect to obtain an index $k_r$ close to the unity if we fit the core-shift only in the conical region.
Thus, we fitted the parabolic and conical regions separately.
In the former, sampled by the frequency range (8-22) GHz, we determine an index of $k_r = 0.57 \pm 0.17$, while in the latter, sampled at 1, 5, and 8 GHz, we obtain $k_r = 0.93 \pm 0.01$.
The point at 8 GHz is used to connect the two regions, since it is the closest one to the jet break.
After the break, we obtain an index slightly lower than the unity, in reasonable agreement with the equipartition hypothesis.
The other two hypotheses are not fully matched.
Indeed, $\alpha = -0.5$ is flatter with respect to the observed values \citep[see][]{Park_2021} and the jet does not show a conical shape up to its apex. 
Nevertheless, we use Eq.\ \ref{eq:B_field} to obtain an order-of-magnitude estimate of the magnetic field strength.
The core-position offset $\Omega_{r \nu}$ is computed between $\nu_2 = 1.4 \, \mathrm{GHz}$ and $\nu_1 = 5.0 \, \mathrm{GHz}$, since these frequencies sample the one-parsec region.
The Doppler factor is estimated to be $\delta = 1.24 \pm 0.25 $ with $\beta = 0.95 \pm 0.03$ (observed at one pc, see Fig.\ \ref{fig:J_CJ_ratio}).
We assume an intrinsic half opening angle of $\phi = (2.5 \pm 1)\degree$ (Fig.\ \ref{fig:opening_angle}). 
Using Eq.\ \ref{eq:B_field} we find $B_1 = 0.13 \pm 0.02 \ \mathrm{G}$.
The magnetic field strength matches the values obtained by \citet{O'Sullivan_2009} in their small samples of objects and it is fairly smaller than the median value $B_1 \sim 0.4 \, \mathrm{G}$ obtained for the BL Lac sources in the MOJAVE sample \citep{Pushkarev_2012}.

\begin{center}
\begin{figure}[t]
    \includegraphics[width=\linewidth]{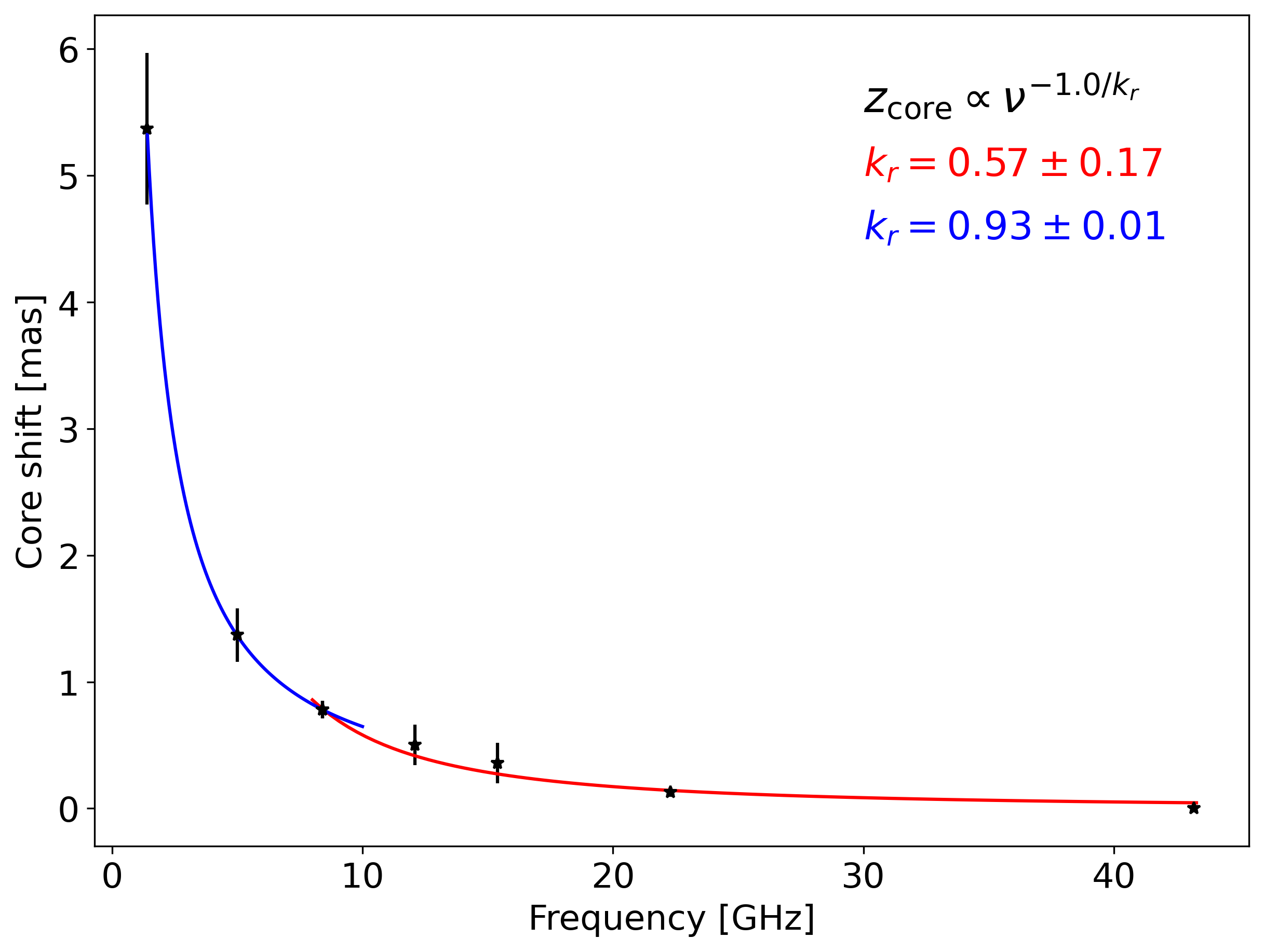}
    \caption{Core positions as a function of frequency, assumed from B21. The data are fitted with two different power laws, one for the parabolic region (red curve) and one for the conical region (blue curve). In the former case, the best fit index is $k_r = 0.57 \pm 0.17$, while in the latter is $k_r = 0.93 \pm 0.01$. An index close to 1 is expected when assuming equipartition between the particles and magnetic field energy densities in the jet.}
    \label{fig:core_shift}
\end{figure}
\end{center} 

In the following, we propose a new formalism to compute the magnetic field for a quasi-parabolic accelerating jet, which better describes the case of NGC\,315. 
We calculate expressions for a magnetic field at a generic distance from the jet base assuming that the flow radius evolves as $r\sim \mathsf{z}^{\psi}$, with $\psi=0.5$ corresponding to a strictly parabolic jet boundary. 
The details are reported in Appendix \ref{app:cs}.
First, we assume the Doppler factor $\delta$ as a constant. By recalculating the position of the surface with optical depth equal to unity, the expression for $\Omega_{r\nu}$ is still Eq.\ \ref{eq:O_rnu} but with $k_r=\psi$, and the magnetic field is:
\begin{equation}
    B_\mathrm{z} = 0.025 \ \bigg(\frac{\Omega_{r \nu}^{3\psi} (1 + z)^2}{\mathsf{z}^{3\psi}r_\mathsf{z}\delta^2 \mathrm{sin}^{3\psi-1}\theta}\bigg)^{1/4} \ \mathrm{G} \, .
    \label{eq:B_field_1}
\end{equation}
Here, $r_\mathsf{z}$ in parsec is the jet radius at $\mathsf{z}$ parsec from the jet base. This expression transforms into Eq.\ \ref{eq:B_field} for $\mathsf{z}=1$, $\psi=1$ and noting that $\varphi\approx r_\mathsf{z}/\mathsf{z}=r_\mathsf{z}$ at one parsec. 

For an accelerating flow we assume a linear acceleration with the jet radius as $\Gamma=r/R_L$  and with the Doppler factor changing along the accelerating flow. 
We distinguish between three possible cases, depending on the interplay between the viewing angle $\theta$ and the Lorentz factor $\Gamma$ in the expression for $\delta$.

Case A): the viewing angle $\theta\sim\Gamma^{-1}$ is such that the Doppler factor is close to its peak, so we may set it as a constant.
In this scenario, $k_r=2\psi$ and the expression for the magnetic field becomes (see details in Appendix~\ref{app:cs})
\begin{equation}
    B_\mathrm{z} = 0.025 \ \Bigg(\frac{\Omega_{r \nu}^{6 \psi} (1 + z)^2}{\mathsf{z}^{6 \psi}r_\mathsf{z}\delta^2 \mathrm{sin}^{6 \psi -1}\theta}\Bigg)^{1/4} \ \mathrm{G} \, .
    \label{eq:B_field_2}
\end{equation}
with $\mathsf{z}$ and $r_\mathsf{z}$ in parsec.

Case B): for a smaller viewing angle $\theta\lesssim\Gamma^{-1}$, we may use 
the approximations $\beta\approx 1-1/(2\Gamma^2)$ and $\cos\theta\approx 1-\theta^2/2$, and by neglecting the factor $\theta^2$ we obtain $\delta\approx 2\Gamma$. The core shift coefficient becomes $k_r=4 \psi/3$, and the magnetic field is given by
\begin{equation}
    B_\mathrm{z} = 0.018 \ \Bigg(\frac{\Omega_{r \nu}^{4 \psi} (1 + z)^2}{\mathsf{z}^{4 \psi}r_\mathsf{z} \mathrm{sin}^{4\psi-1}\theta}\left(\frac{R_L}{r_\mathsf{z}}\right)^2\Bigg)^{1/4} \ \mathrm{G}
    \label{eq:B_field_3}
\end{equation}
with the light cylinder radius $R_L$ measured in parsec.

Case C): for a larger viewing angle $\theta\gtrsim\Gamma^{-1}$, we use the exact expression for a Doppler factor with variable $\Gamma$ but assuming $\beta=1$, since the term $\beta\cos\theta$ in $\delta$ is not close to unity, and the difference of velocity from speed of light is not important. In this case $k_r=8 \psi/3$, and the magnetic field is expressed as
\begin{equation}
    B_\mathrm{z} = 0.025 \ \Bigg(\frac{\Omega_{r \nu}^{8 \psi} (1 + z)^2 (1-\cos\theta)^2}{\mathsf{z}^{8 \psi}r_\mathsf{z} \mathrm{sin}^{8 \psi-1}\theta}\left(\frac{r_\mathsf{z}}{R_L}\right)^2\Bigg)^{1/4} \ \mathrm{G} \, .
    \label{eq:B_field_4}
\end{equation}
For NGC\,315, the viewing angle is $\theta \sim 0.66 \, \mathrm{rad}$ and the inverse of the Lorentz factor evolves with distance from $\Gamma^{-1} \sim 1$ at the jet base to $\Gamma^{-1} \sim 0.5$ at $\mathsf{z}_\mathrm{br}$.
Therefore, Case A) is the most adequate at the transition distance,  while Case B) is true in the accelerating part of the jet.
At $\mathsf{z}= \mathsf{z}_\mathrm{br} =  0.58 \pm 0.28 \, \mathrm{pc}$, we obtain $B_{\mathsf{z}_\mathrm{br}} = 0.18 \pm 0.06 \, \mathrm{G}$ and $B_{\mathsf{z}_\mathrm{br}} = 0.013 \pm 0.003 \, \mathrm{G}$, respectively.
As we expect a larger magnetic field value closer to the jet base, the estimate in the Case A) corresponds better to the value at $\mathsf{z}=1$~pc than in the Case B). 
However, we note that Eq.\ \ref{eq:B_field_3} contains a factor $(R_L/r_\mathsf{z})^{2}$ that follows from the ideal linear acceleration $\Gamma=r/R_L$. 
Slower accelerations or the maximum Lorentz factor reached closer to the jet axis will boost this magnetic field estimate \citep{Beskin_2017}.

We note, that the relation $k_r=4 \psi /3$, valid in Case B), is in excellent agreement with the observations, being $\psi=0.45$ and the measured $k_r$ at high frequencies $k_r=0.57\pm 0.17$. Instead, the relation $k_r=2 \psi$, valid in Case A), implies a $k_r$ value beyond the observed interval.
 
\subsection{ 
Magnetic field: from the parsec jet up to the accretion disk}
\label{sec:B_field_accretion}

In this Section we explore the properties of the magnetic fields in the nuclear region in two different ways: 1) from the disk magnetic flux (Sect.\ \ref{sec:BH_spin}) assuming a simplified geometry for the disk and 2) by extrapolating up to the jet base our field strengths estimation (Sect.\ \ref{sec:B_field}).
The results are compared with the expected field strength needed to saturate the disk and form a MAD, whose presence is suggested from the comparison of the magnetic fluxes (Sect.\ \ref{sec:BH_spin}).

\begin{itemize}
    \item Saturation magnetic field
\end{itemize}
In the MAD state, the accretion disk is disrupted by the accumulated poloidal field within the extent of the magnetosphere ($r_\mathrm{m}$ in units of $R_\mathrm{S}$), which corresponds to the jet launching radius.
The magnetosphere radius, following \citet{Narayan_2003}, is defined as:
\begin{equation}
    r_\mathrm{m} \sim 8.2 \times 10^3 (\epsilon \times 10^2)^{2/3} m_8^{-2} \dot{m}^{-2/3} \bigg(\frac{\Phi}{0.1 \, \mathrm{pc^2} \, \mathrm{G}}\bigg)^{4/3} \, R_\mathrm{s}
    \label{eq:magnetosphere_radius}
\end{equation}
where:
\begin{itemize}
    \item $\epsilon$ defines the ratio between the radial velocity of the gas within the magnetosphere and its free-fall velocity. We assume an upper limit of 0.1, following \citet{Narayan_2003};
    \item $m_8 = M_{\mathrm{BH}} / (10^8 M_\odot)$;
    \item $\dot{m} = \dot{M} / \dot{M_\mathrm{Edd}}$;
    \item $\Phi$ is the magnetic flux in the central region, for which we adopt our $\Phi$ estimation with $b=2$.
\end{itemize}
We infer the upper limit of the magnetosphere radius to be $r_m = 1.1 \pm 1.0 \, R_\mathrm{S}$.
The error is computed by propagating the uncertainties on the magnetic flux and the accretion mass rate.
We notice that in NGC\,315 the event horizon radius varies between $r_\mathrm{H} \in [0.6,0.8] \, R_\mathrm{s}$ when $a_* = 0.99$ and $a_* = 0.86$, respectively.
Since a magnetosphere smaller than an event horizon would be unphysical, the lower limit on $r_m$ is set by the lower limit on the event horizon radius.
Therefore, the magnetosphere radius lays in the range $\sim 0.6 - 2.1 \, R_\mathrm{S}$, or $\sim 1.2 - 4.2 \, r_\mathrm{g}$ when $\eta_d = 0.1$ and $\sim 0.6 - 5.3 \, R_\mathrm{S}$, or $\sim 1.2 - 10.6 \, r_\mathrm{g}$ when $\eta_d = 0.4$.
The derived range is relatively small with respect to estimates from simulations. For instance, \citet{Narayan_2021} inferred magnetosphere radii on the order of several tens of $r_\mathrm{g}$.

Next, we extrapolate the saturation magnetic field in the accretion disk by knowing that a MAD forms when enough field is accumulated in it, so that the magnetic force  $F_\mathrm{B}$ overcomes the gravitational force $F_\mathrm{G}$.
When $F_\mathrm{B}  > F_\mathrm{G}$, the magnetic field is expelled from the disk until the condition $F_\mathrm{B} = F_\mathrm{G}$ is reached.
Following \citet{Contopoulos_book}, we express the forces as $F_\mathrm{B} = \big(B^2/ 8 \pi \big) 4 \pi r^2 h/r$ and $F_\mathrm{G} = G M_\mathrm{BH} M_\mathrm{disk} / r^2$, in which $h/r$ is the scale height of the accretion disk, $r = r_\mathrm{m}$ is the radius and $M_\mathrm{disk}$ is the disk mass in the region in which the MAD forms. 
Writing the disk mass as $M_\mathrm{disk} = 4/3 \rho \pi r^3 h/r$ and the density $\rho$ from the mass continuity equation as $\rho = \dot{M} / \big(4 \pi r^2\ v_r h/r \big)$ yields to a saturation magnetic field:
\begin{equation}
    B_\mathrm{MAD} = \bigg(\frac{2}{3} \frac{G M_\mathrm{BH} \dot{M}}{r^3 v_r h/r}\bigg)^{1/2}
    \label{eq:B_MAD}
\end{equation}
in which $v_r$ is the radial velocity of the infalling gas and is assumed to be close to the speed of light.
We solve Eq.\ \ref{eq:B_MAD} by adopting the mass accretion rate derived from the [O III] line luminosity in Sect.\ \ref{sec:BH_spin}.
We set the scale height to $0.25 \pm 0.10$, accordingly to \citet{Narayan_2021} (see Fig.\ 7, right lower panel in that article) and extrapolate a saturation magnetic field spanning the range $B_\mathrm{MAD} = 125 - 480 \, \mathrm{G}$ at $r_\mathrm{m} = 0.6 \, R_\mathrm{S}$ and $B_\mathrm{MAD} = 5 - 18 \, \mathrm{G}$ at $r_\mathrm{m} = 5.3 \, R_\mathrm{S}$.

\begin{itemize}
    \item Poloidal field from the magnetic flux $\Phi$
\end{itemize}
In the accretion disk within the supposed magnetosphere, the strength of the poloidal magnetic field component $B_P$ is related to the magnetic flux $\Phi$ in the flux freezing approximation.
The poloidal strength is $B_P = \Phi/S$, where we assume a circular surface $S = \pi r^2$ with $r \in [0.6,5.3]R_\mathrm{S}$.
The field varies between $B_P = 4640-464 \, \mathrm{G}$ and $B_P = 60-6 \, \mathrm{G}$ at the lower and upper limit of the magnetosphere.

\begin{itemize}
    \item Extrapolation from the field strength in the jet 
\end{itemize}
Finally, we extrapolate the magnetic field values inferred in Sect.\ \ref{sec:B_field} up to the jet base.
To do that, we assume that the magnetic flux is conserved along the sub-parsec jet.
Under this assumption, both the toroidal and poloidal components of the magnetic field in the parabolic region evolve as
\begin{equation}
B(\mathsf{z}) = B_0 (\mathsf{z}_0 / \mathsf{z})^{2\psi} \, .
\end{equation}
The details are shown in Appendix \ref{app:B_field_evol}.
Being $\psi = 0.45$ upstream of the transition distance, both the poloidal and toroidal magnetic field components evolve as $\sim \mathsf{z}^{-1}$, while in the conical region the poloidal component evolves as $\sim \mathsf{z}^{-2}$.

\begin{center}
\begin{figure}[t]
    \includegraphics[width=\linewidth]{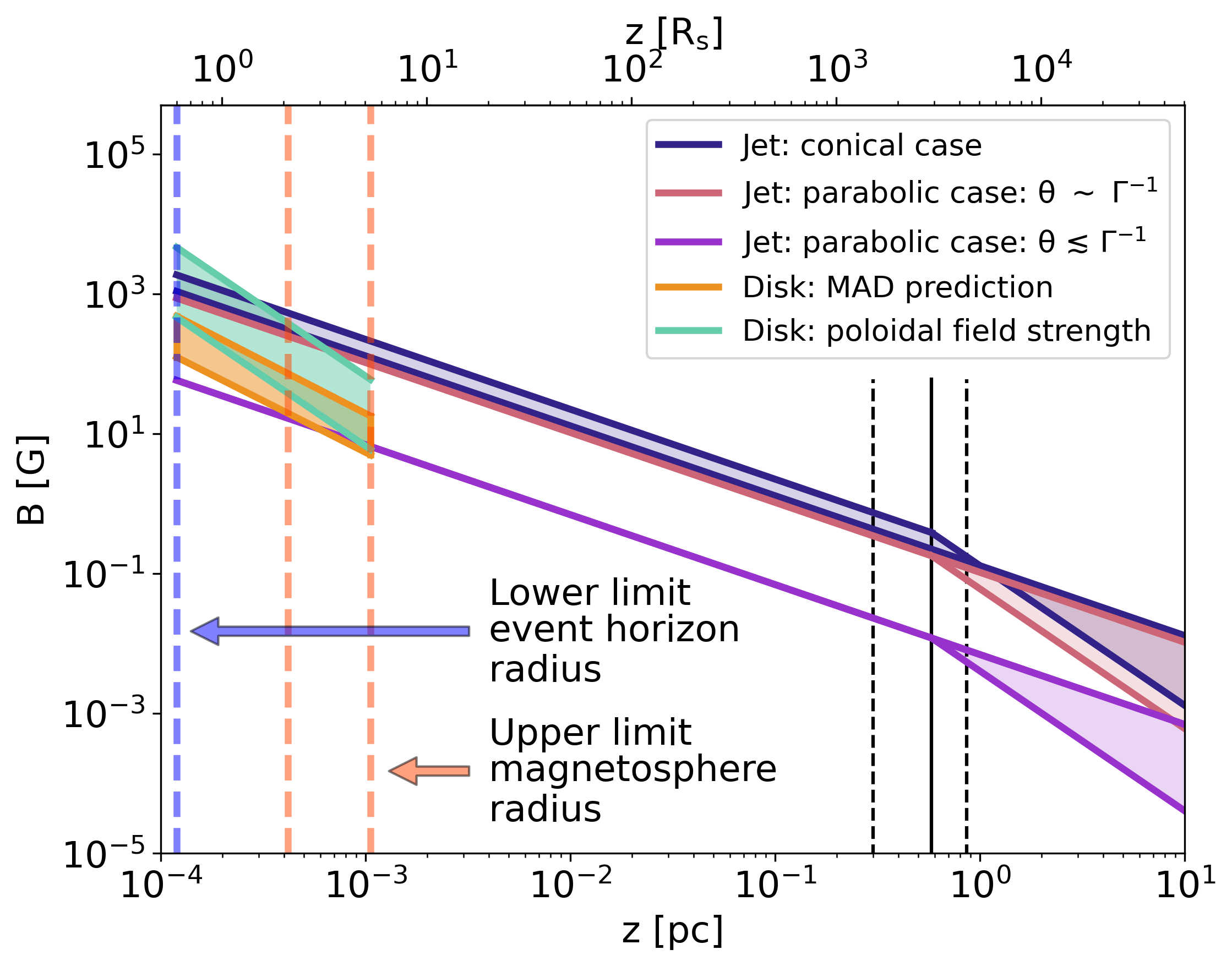}
    \caption{Extrapolation of the magnetic field computed using Eq.\ \ref{eq:B_field} (blue lines), Case A) (red lines), and case B) (purple lines) down to the event horizon radius $r_\mathrm{H} = 0.6 \, R_\mathrm{S}$ (highlighted by the blue dashed line) assuming $B(\mathsf{z}) = B_\mathrm{0} (\mathsf{z} / \mathsf{z}_0)^{2\psi}$. See the text for the details on the different dependence. The orange dashed lines indicate the upper limit on the magnetosphere radius, namely $r_m = 2.1 \, R_\mathrm{S}$ with $\eta_d = 0.1$ and $r_m = 5.3 \, R_\mathrm{S}$ with $\eta_d = 0.4$. The orange box highlights the range of variation of the expected saturation magnetic field strength in the MAD scenario, while the light green box indicates the poloidal field strengths inferred based on the magnetic flux.}
    \label{fig:B_profile}
\end{figure}
\end{center}

Our results are summarized in Fig.\ \ref{fig:B_profile}. 
From the first two methods, the magnetic field strength in the accretion disk is found to be in the range of $\sim 10-10^3 \, \mathrm{G}$, based on the possible extent of the magnetosphere.
On the other hand, the strengths at the jet base are up to tens of Gauss by extrapolating Case B) and up to $\sim 10^3 \, \mathrm{G}$ by extrapolating Case A) and the conical case.
A comparable range ($360 \, \mathrm{G} < \mathrm{B} < 6.9 \times 10^4 \, \mathrm{G}$ at 1 $R_\mathrm{S}$ from the core) was derived by \citet{Baczko_2016} in the radio galaxy NGC\,1052.

From General Relativistic MagnetoHydrodynamical (GRMHD) simulations, the magnetic energy ($\propto B^2$) in a Blandford-Znajek jet is expected to be one-two orders of magnitude stronger than the one in the disk \citep{McKinney_2007a,McKinney_2007b}.
In this scenario, the conical and A) cases are favoured leading to fields orders of magnitude stronger than the strengths needed to form a magnetically arrested disk, especially at the largest magnetosphere radius.

\subsection{Jet power: expected vs. observed} \label{sec:jet_power}
A final test of the models described above can be performed by comparing the expected jet power to the observed one.
By knowing the accretion rate, the expected jet power can be computed
as $P_\mathrm{jet} = \eta \dot{M} c^2$.
By means of GRMHD simulations within the MAD scenario,  \citet{Narayan_2021} linked different black hole spins to the expected accretion efficiencies.
From our determined spin range $a_*=0.72-0.99$ we expect $\eta \in [0.72, 1.50]$ in the prograde case (i.e.\ the black hole and the accretion disk rotate in the same direction) and $\eta \in [0.14, 0.42]$ in the retrograde case (i.e.\ the black hole and the accretion disk rotate in opposite directions).
Using the [OIII] line estimation for the accretion rate $\dot{M}$ and $\eta_d = 0.1$, the jet power of NGC\,315 is $P_\mathrm{jet} \sim 1 - 4 \times 10^{44} \, \mathrm{erg/s}$ in the prograde case and $P_\mathrm{jet} \sim 0.3 - 1 \times 10^{44} \, \mathrm{erg/s}$ in the retrograde case.
In the case of $\eta_d = 0.4$, the jet power is expected to be $P_\mathrm{jet} \sim 4 - 9 \times 10^{43} \, \mathrm{erg/s}$ and $P_\mathrm{jet} \sim 0.9 - 3 \times 10^{43} \, \mathrm{erg/s}$, respectively

A further method to compute the total jet power is provided by \citet{Nohkrina_2019} based on their model (see Sect.\ \ref{sec:BH_spin}):
\begin{equation}
    P_\mathrm{jet} = \frac{c}{8} \bigg(\frac{\Phi}{\pi R_\mathrm{L}} \bigg)^2 \, .
    \label{eq:jet_Nokhrina}
\end{equation}
By assuming $b = 1$ and $b = 2$ we obtain respectively the lower and upper limit for the jet power, $P_\mathrm{jet} \sim 2 \times 10^{41} \, \mathrm{erg} \, \mathrm{s^{-1}}$ and $P_\mathrm{jet} \sim 1 \times 10^{44} \, \mathrm{erg} \, \mathrm{s^{-1}}$.
The comparison with the previous $P_\mathrm{jet}$ estimates favours an index for the external pressure $b \sim 2$, consistent with our findings in Sect.\ \ref{sec:BH_spin}.

In the following, we compare these expected values with observational estimates of the jet power in NGC\,315.
We first work under the assumption of a Bondi accretion flow \citep{Bondi_1952}.
Following \citet{Russell_2015}, we can derive the accretion rate, in convenient units, as:
\begin{equation}
    \dot{M}_\mathrm{B} = 0.012 \, \bigg(\frac{k_\mathrm{B}T}{\mathrm{keV}}\bigg)^{-3/2} \, \bigg(\frac{n_e}{\mathrm{cm^{-3}}}\bigg) \, \bigg(\frac{M_\mathrm{BH}}{\mathrm{10^9 \, M_\odot}}\bigg)^2 \, \mathrm{M_\odot \, yr^{-1}}
    \label{eq:Bondi_acc}
\end{equation}
in which $n_e$ is the electron number density and $k_\mathrm{B}T$ is the temperature of the medium at the Bondi radius.
In this equation, an adiabatic index of 5/3 is assumed.
The parameters of the gas within the Bondi radius, $k_\mathrm{B}T = 0.44 \, \mathrm{keV}$ and $n_e = 2.8 \times 10^{-1} \, \mathrm{cm^{-3}}$, are extracted from X-ray Chandra data \citep{Worrall_2007}.
Eq.\ \ref{eq:Bondi_acc} yields an accretion rate of $\dot{M}_\mathrm{B} = 3.1 \times 10^{24} \, \mathrm{g/s} = 5 \times 10^{-2} \, \mathrm{M_\odot / yr}$, around 100 times lower than the Eddington accretion rate $M_\mathrm{Edd} = 2.9 \times 10^{26} \, \mathrm{g/s}$.
\citet{Nemmen_2015} and \citet{Allen_2006}, among others, observed that a relation exists between the Bondi accretion rate and the jet power.
The latter can be directly inferred as $\mathrm{log} P_\mathrm{jet} = A \, \mathrm{log}(0.1 \dot{M}_\mathrm{B}c^2) + B$ in which the parameters are either $A = 1.05$ and $B = -3.51$ \citep{Nemmen_2015} or $A = 1.29$ and $B = - 0.84$ \citep{Allen_2006} meaning $P_\mathrm{jet} \sim 2 \times 10^{43} \, \mathrm{ erg \, s^{-1}}$ and $P_\mathrm{jet} \sim 1 \times 10^{44} \, \mathrm{ erg \, s^{-1}}$, respectively. 
A second method relies on the existence of an empirical relationship between the radio core luminosity and the jet power. Following the work of \citet{Heinz_2007}, \citet{Morganti_2009} inferred $P_\mathrm{jet} = 1.4 \times 10^{44} \, \mathrm{ erg \, s^{-1}}$. While the validity of such a relation may, in general, be affected by the core variability and Doppler boosting, we do not expect these effects to be strong in NGC\,315, and the obtained value lies in the same range as the previous estimates. 

This analysis shows a substantial agreement between the order-of-magnitude of the jet powers computed in the MAD scenario and from Eq.\ \ref{eq:jet_Nokhrina} and the observational estimates.
Obtaining tighter constraints concerning, for instance, which spin configuration is the most likely is not possible based on the current data due to the large uncertainties and the degeneracy between the different parameters. 

\section{Conclusion} \label{sec:Conclusions}

In this paper we have explored the disk-jet connection in the nearby radio galaxy NGC\,315 using a multi-frequency and multi-epoch VLBI data set.
We constrained the observational properties of the jet in the acceleration and collimation region and used them to retrieve information on the physics of the accretion disk and of the jet itself.
Our results are here summarized.
\begin{itemize}
    \item We used simultaneous 22 GHz and 43 GHz observations to study the evolution of the spectral index on VLBI scales. 
    We inferred an unexpected spectral behavior, with very steep values ($\alpha \sim -1.5$) in a region whose extent is comparable with that of the collimation zone.
    These steep values may result from strong synchrotron losses at the highest frequencies, reflecting the presence of strong magnetic fields at the jet base.
    \item Using a pixel-based analysis, we derived the jet speed profile from the observed jet-to-counterjet ratio profile, under the assumption of intrinsic symmetry between the two jets. We inferred a fast-acceleration on sub-parsec scales, where the jet collimation is also taking place. The co-spatiality between the two phenomena suggests the jet to consist of a cold outflow in which the acceleration is mainly driven by the conversion of magnetic energy into kinetic energy of the bulk.
    Following this, we modeled the jet acceleration in the context of MHD theories.
    \item We found the half intrinsic opening angle to span the range between $\sim 1\degree$ and $\sim 6\degree$, consistent with typical values observed in radio galaxies.
    The opening angle decreases on sub-parsec scales down to approximately constant values of $\sim 2\degree-3\degree$.
    Consequently, we examined the $\Gamma \phi$ product along the jet, finding a fairly constant profile that varies between 0.04 and 0.2, with $\Gamma \phi = 0.07$ as average.
    This value, is consistent with the median product of $\Gamma \phi = 0.13 - 0.17$ from sample studies.
    At the jet base, we indirectly estimate the average $\Gamma \phi$ product to be likely close to the expectation of $\Gamma \phi \sim 1$ for magnetically accelerating jets. The discrepancy between the expected and the observed value may be explained if either the $\Gamma \phi$ product drops fast in the acceleration region or if the condition $\Gamma \phi \sim 1$ is only valid in  the very central jet spine, and the observed values mainly reflect the properties of the outer sheath. 
    \item Under the assumption that an equipartition state is reached at the end of the acceleration and collimation region, and in the framework of the model proposed by \citet{Nohkrina_2019, Nokhrina_2020}, we found the black hole of NGC\,315 to be fast rotating, with $a_* \gtrsim 0.72$.  Moreover, based on the X-ray properties of the accretion flow, we constrained the total magnetic flux threading the accretion disk in to be in excellent agreement with the expected flux in case of a magnetically arrested disk.
    The two are estimated independently, with the former being extrapolated from the jet properties at the transition and the expected one based on the accretion rate and black hole mass.
    The average dimensionless magnetic flux $\phi_\mathrm{BH} \sim 40$ is in a good agreement with the saturation value of $\sim 50$, expected when the accretion disk reaches the magnetically arrested state. 
    \item We estimated three different magnetic field values down to the jet base based on the core-shift effect in the case of both a conical and a quasi-parabolic accelerating jet.
    The latter case required the development of a new formalism.
    In the case of $\theta \sim \Gamma^{-1}$, we found a magnetic field strength of $B \sim 0.18 \, \mathrm{G}$ at $\mathsf{z}_\mathrm{br} = 0.58 \, \mathrm{pc}$ consistent with the one in the conical assumption, while if $\theta \lesssim \Gamma^{-1}$ the magnetic strength is lower, $B \sim 0.01 \, \mathrm{G}$.
    Later, we extrapolated the field values down the event horizon radius assuming both a poloidal and toroidal field configuration.
    The extrapolated strengths are consistent with both the poloidal field values obtained from the magnetic disk flux and the needed saturation field strengths to form a MAD.
    This result provides evidence that the magnetic field in the accretion disk has reached high enough values to saturate, establishing a magnetically arrested state.
\end{itemize}

In conclusion, our analysis and modeling  of the jet observational properties on VLBI scales are compatible with the expectations from an accretion disk that has reached a magnetically arrested state.
Our results will be expanded in future papers by means of numerical simulations and new VLBI data sets, especially multi-epoch 86 GHz GMVA observations.

\begin{acknowledgements} 
We thank the referee, Richard A. Perley, for the insightful comments that helped improve the overall clarity of the manuscript.
We would like to thank Maciek Wielgus, for his helpful comments.
L.R., B.B., and E.M. acknowledge the financial support of a Otto Hahn research group from the Max Planck Society.
E.E.N. was supported by the Russian Science Foundation (project 20-62-46021).
M.P. acknowledges support by the Spanish Ministerio de Ciencia through grants PID2019-105510GB-C31 and PID2019-107427GB-C33, and from the Generalitat Valenciana through grant PROMETEU/2019/071.
This research has made use of data from the MOJAVE database that is maintained by the MOJAVE team \citep{Lister_2018}. 
The VLBA is a facility of the National Science Foundation under cooperative agree- ment by Associated Universities, Inc.
The European VLBI Network is a joint facility of European, Chinese, South African and other radio astronomy institutes funded by their national research councils.

\end{acknowledgements}

\bibliographystyle{aa}
\bibliography{bibliography.bib}

\appendix

\section{Core-shift in a quasi-parabolic flow} \label{app:cs}

We suppose that the flow has a shape $r\sim \mathsf{z}^{\psi}$, with $\psi=0.5$ corresponding to a strictly parabolic jet boundary. 
We start with Eq.\ 6 in \citet{Nokhrina_2015}:
\begin{equation}
\left(\nu\frac{1+z}{\delta}\right)^{2.5-\alpha}=c(\alpha)\frac{1+z}{\delta}\frac{2r}{\sin\theta}r_0^2\nu_0\left(\frac{e}{2\pi mc}\right)^{1.5-\alpha}k_eB^{1.5-\alpha}.
\label{app:init}
\end{equation}
Here we assume the following emitting particle energy distribution $dn=k_e\gamma^{-1+2\alpha}d\gamma$, $\gamma\in[\gamma_{\rm min};\,\gamma_{\rm max}]$ where $r_0$ is the electron classical radius, $\nu_0=c/r_0$, the function $c(\alpha)$ is defined in \citet{Hirotani_2005}, $n$, $k_e$ and $B$ are in the plasma proper frame, while all the rest of parameters ($\nu$, $\theta$, etc.) are in the observer's frame. 
The standard assumption of equipartition between magnetic field and emitting plasma energy density \citep{Blandford_1979, Lobanov_1998, Hirotani_2005} is written as
\begin{equation}
k_e=\frac{B^2}{8\pi mc^2\Lambda},\label{app:eqiu}
\end{equation}
where $\Lambda=\ln\frac{\gamma_{\rm min}}{\gamma_{\rm max}}$ for $\alpha=-1/2$, and $\Lambda=\left(\gamma_{\rm max}^{1+2\alpha}-\gamma_{\rm min}^{1+2\alpha}\right)/(1+2\alpha)$ for $\alpha\ne -1/2$. 
The geometrical depth along the line of sight in the conical geometry is equal to $2\mathsf{z}\phi/\sin{\theta}$ in the nucleus frame while for a quasi-parabolic jet is $2r/\sin{\theta}$, where $r=a\mathsf{z}^{\psi}$ is the jet radius in the core at any given frequency.

First, we suppose that the jet has a constant Lorentz factor. 
We use the following dependencies for the particle number density and magnetic field in the plasma proper frame $n\propto r^{-2}\propto\mathsf{z}^{-2\psi}$ and $B\propto r^{-1}\propto\mathsf{z}^{-\psi}$. 
The first expression indicates the continuity of a flow with a constant speed while the second one corresponds to a magnetic field created by an electric current flowing along the jet axis. 
The Doppler factor $\delta$ is a constant and only changes in derivation of the expressions for $\Omega_{r\nu}$, while $B$ is connected with the different geometry.
Setting the constant $C=2c(\alpha)r_0^2\nu(e/2\pi mc)^{1.5-\alpha}$ we obtain
\begin{equation}
\nu^{2.5-\alpha}=C\left(\frac{\delta}{1+z}\right)^{1.5-\alpha}\frac{r_\mathsf{z}}{\sin\theta}k_{e,\,\mathsf{z}}B_{\mathsf{z}}^{1.5-\alpha}\left(\frac{\mathsf{z}}{Z}\right)^{(2.5-\alpha)\psi},
\label{app:nu1}
\end{equation}
where $k_{e,\,\mathsf{z}}$ and $B_{\mathsf{z}}$ are taken at a distance $\mathsf{z}$, and $Z$ is a position of the core observed at the frequency $\nu$. 
We see that $\nu\propto Z^{-\psi}$ and from Eq.\ \ref{app:nu1} we obtain the same equation for $\Omega_{r\nu}$ as Eq.\ \ref{eq:O_rnu} but with $k_r=\psi$. 
Substituting Eq.\ \ref{app:eqiu} into the relation between $\Omega_{r\nu}$ and the flow physical properties
\begin{equation}
C\left(\frac{\delta}{1+z}\right)^{1.5-\alpha}\frac{r_\mathsf{z}}{\sin\theta}k_{e,\,\mathsf{z}}B_{\mathsf{z}}^{1.5-\alpha}=\left(\frac{\Omega_{r\nu}}{\mathsf{z}\sin\theta}\right)^{(2.5-\alpha)\psi},
\end{equation}
we obtain Eq.\ \ref{eq:B_field_1} for a magnetic field for $\alpha=-1/2$.

For the accelerating flow we assume linear acceleration with a jet radius $\Gamma=r/R_L$. 
This means that the dependencies for a particle number density and a magnetic field are different. 
Indeed, the flow continuity leads in the plasma proper frame to $n\propto r^{-2}\Gamma^{-1}\propto\mathsf{z}^{-3\psi}$ while the toroidal magnetic field, created by an electric current, in the plasma proper frame can be written as $B\propto r^{-1}\Gamma^{-1}\propto \mathsf{z}^{-2\psi}$. 
In this case, if there is equipartition at some distance $\mathsf{z}_0$, it does not hold everywhere in a jet. 
The emitting particle energy starts slowly dominating downstream until the acceleration becomes very slow. 
However, for a mildly relativistic flow the departure from equipartition, which is of the order $u_{\rm part}/u_{\rm B}\approx \Gamma/\Gamma_0$ with $\Gamma_0$ being the bulk motion Lorentz factor at $\mathsf{z}_0$, is not dramatic. 
The second difference comes from the Doppler factor changing along the accelerating flow. 
For a given observational angle $\theta$, the Doppler factor has a maximum at approximately $\cos\theta=\beta$, which for a small angle $\theta\ll 1$ transforms into $\Gamma\approx\theta^{-1}$. 
For NGC\,315 ($\theta=38\degree$) the Doppler factor has a maximum at $\Gamma\approx 1.6$ and can be treated approximately as a constant. 
For the smaller Lorentz factors $\delta\sim 2\Gamma$, while for the larger $\delta\sim 1/\Gamma(1-\cos\theta)$. 
Thus, we consider three cases for an accelerating flow depending on the value of a Lorentz factor at the cores. 

In case A) the Doppler factor is close to its peak and the emission from the fastest parts of a flow reaches the observer as $\theta\sim\Gamma^{-1}$. 
Substituting the dependencies for $n$, $B$ and $r$ on $Z$ into Eq.\ \ref{app:init}, we obtain
\begin{equation}
\nu^{2.5-\alpha}=C\left(\frac{\delta}{1+z}\right)^{1.5-\alpha}\frac{r_\mathsf{z}}{\sin\theta}k_{e,\,\mathsf{z}}B_{\mathsf{z}}^{1.5-\alpha}\left(\frac{\mathsf{z}}{Z}\right)^{(5-2\alpha)\psi}.
\label{app:nu2}
\end{equation}
From this, Eq.\ \ref{eq:O_rnu} for $\Omega_{r\nu}$ with $k_r=2\psi$ follows. 
Now, the relation of the core shift with jet physical parameters is
\begin{equation}
C\left(\frac{\delta}{1+z}\right)^{1.5-\alpha}\frac{r_\mathsf{z}}{\sin\theta}k_{e,\,\mathsf{z}}B_{\mathsf{z}}^{1.5-\alpha}=\left(\frac{\Omega_{r\nu}}{\mathsf{z}\sin\theta}\right)^{(5-2\alpha)\psi},
\end{equation}
and assuming a close-to-equipartition state we obtain Eq.\ \ref{eq:B_field_2} for the magnetic field.

In case B) the fastest components are observed well within the emission cone as $\theta<\Gamma^{-1}$. 
We regard the case of $\delta=2\Gamma$ and substitute
\begin{equation}
\delta=2\frac{r_\mathsf{z}}{R_{\rm L}}\left(\frac{Z}{\mathsf{z}}\right)^{\psi}
\end{equation}
into Eq.\ \ref{app:init}. 
The dependence of the observational frequency on a distance $Z$ can be rewritten as
\begin{equation}
\nu^{2.5-\alpha}=C\left(\frac{2}{1+z}\frac{r_\mathsf{z}}{R_{\rm L}}\right)^{1.5-\alpha}\frac{r_\mathsf{z}}{\sin\theta}k_{e,\,\mathsf{z}}B_{\mathsf{z}}^{1.5-\alpha}\left(\frac{\mathsf{z}}{Z}\right)^{(3.5-\alpha)\psi}.
\label{app:nu3}
\end{equation}
From this, the relation $k_r=\psi(3.5-\alpha)/(2.5-\alpha)=4\psi/3$ for $\alpha=-1/2$ and Eq.\ \ref{eq:B_field_2} follow.

In case C) emission from the fastest components are de-boosted due to the relation $\theta>\Gamma^{-1}$, so it is of academic interest only. 
We substitute with $\beta\approx 1$
\begin{equation}
\delta=\frac{R_{\rm L}}{r_\mathsf{z}}\left(\frac{\mathsf{z}}{Z}\right)^{\psi}\frac{1}{1-\cos\theta}
\end{equation}
into Eq.\ \ref{app:init} and obtain
\begin{equation}
\begin{array}{rcl}
\displaystyle\nu^{2.5-\alpha}&=&\displaystyle C\left(\frac{2}{1+z}\frac{R_{\rm L}}{r_\mathsf{z}}\right)^{1.5-\alpha}\frac{r_\mathsf{z}}{\sin\theta(1-\cos\theta)}\times\\ \ \\
\displaystyle&\times& \displaystyle k_{e,\,\mathsf{z}}B_{\mathsf{z}}^{1.5-\alpha}\left(\frac{\mathsf{z}}{Z}\right)^{(6.5-3\alpha)\psi}.
\end{array}
\label{app:nu4}
\end{equation}
The corresponding index is $k_r=\psi(6.5-3\alpha)/(2.5-\alpha)=8\psi/3$ for $\alpha=-1/2$. 

\section{Toroidal and poloidal magnetic field extrapolation in an accelerating jet} \label{app:B_field_evol}

The toroidal and poloidal magnetic field components in the core frame (indicated as $'$) are $B_\varphi = B_\varphi'/\Gamma$ and $B_p = B_p'$, respectively.
In Sect.\ \ref{sec:B_field} we computed the magnetic field in the plasma proper frame and we called it $B_z$ for simplicity.
If the toroidal field dominates, $B_z = B_\varphi$, otherwise $B_z = B_p$.
In a relativistic flow, the toroidal and poloidal components are related as $B_\varphi' = B_p' \frac{r}{R_\mathrm{L}}$.
We assume the magnetic flux is conserved along the jet, with a pure poloidal field following $B_p' r^2 = \mathrm{const}$, and we note that if the poloidal component depends on r the relation still holds with some numerical coefficient.
If the toroidal component dominates in the core region, the magnetic field is $B_g = B_p' (r/r_g)^2 = B_\varphi \Gamma R_\mathrm{L}/r (r/r_g)^2$.
In the accelerating region $\Gamma R_\mathrm{L}/r \sim 1$, thus with $r \propto \mathsf{z}^\psi$:
\begin{equation}
    B_g = B_\mathsf{z} \left( \frac{\mathsf{z}}{\mathsf{z}_g} \right)^{2\psi} \, .
\end{equation}
Instead, if the poloidal components dominates in the core region, $B_g = B_p' (r/r_g)^2 = B_\mathsf{z} (r/r_g)^2$, from which 
\begin{equation}
    B_g = B_\mathsf{z} \left( \frac{\mathsf{z}}{\mathsf{z}_g} \right)^{2\psi} \, .
\end{equation}

\section{NGC\,315 images} \label{app:MOJAVE_image}

\begin{figure*}[h]
\centering
    \includegraphics[width=6.3cm]{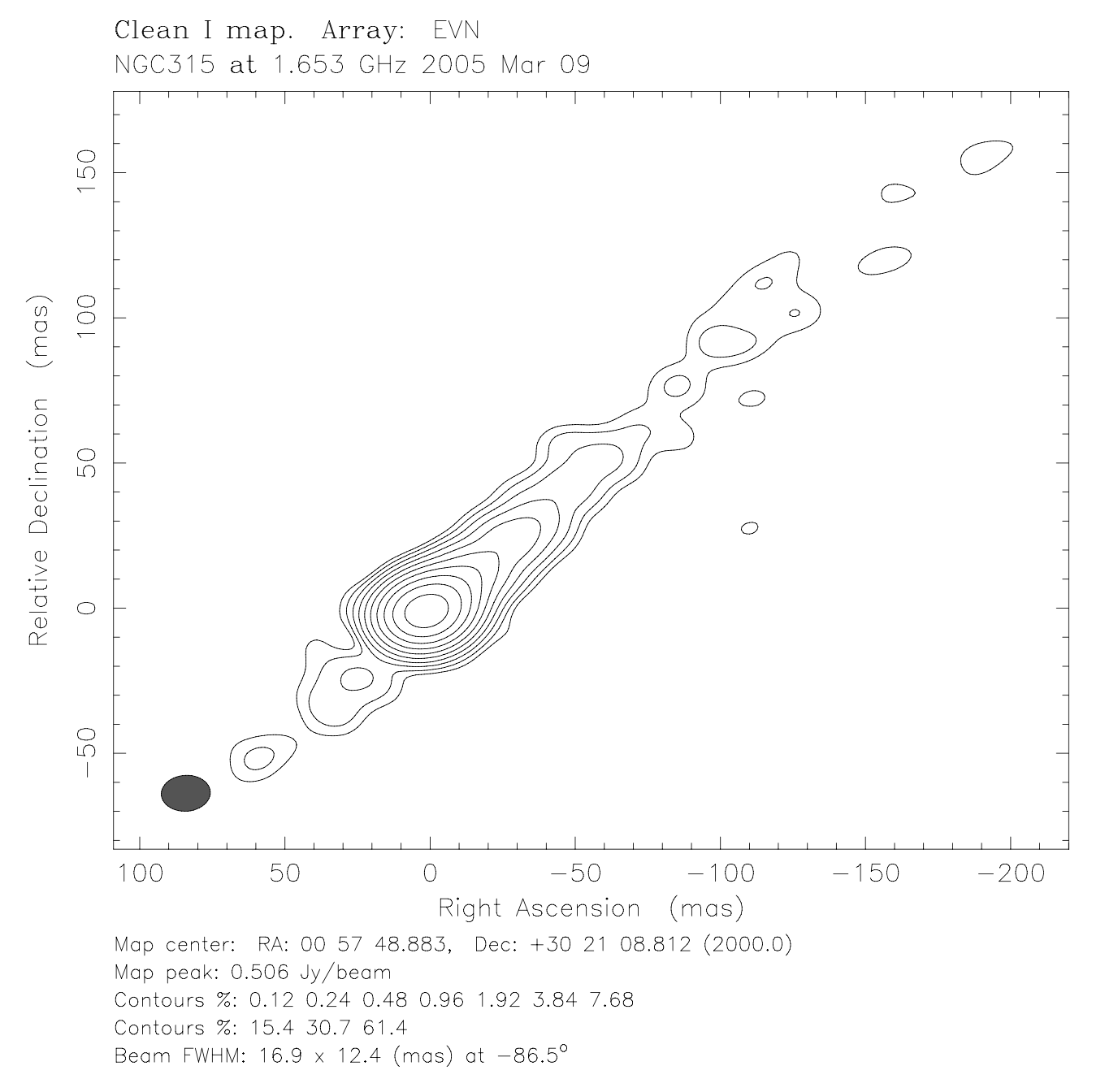}
  \caption{VLBI image of NGC\,315 at 1 GHz.}
  \label{fig:1GHz}
\end{figure*}

\begin{figure*}[h]
\centering
\begin{multicols}{2}
    \includegraphics[width=6.3cm]{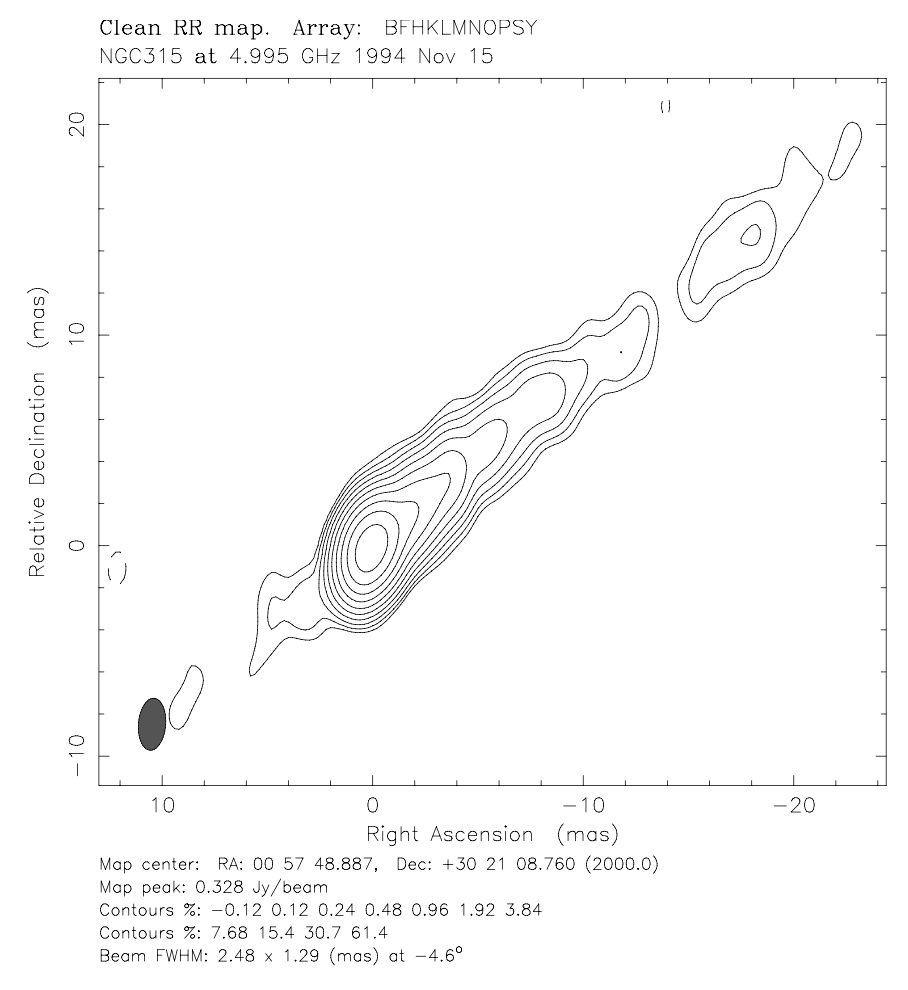}\par
    \includegraphics[width=6.3cm]{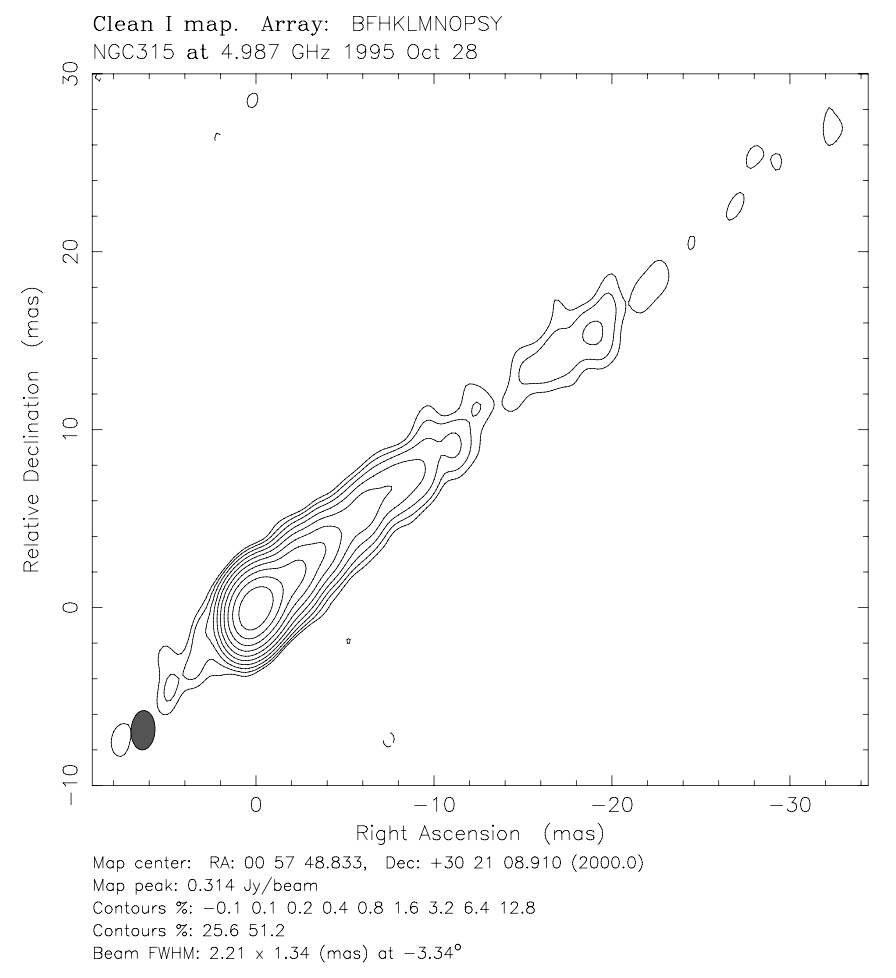}\par
\end{multicols}
\begin{multicols}{2}
    \includegraphics[width=6.3cm]{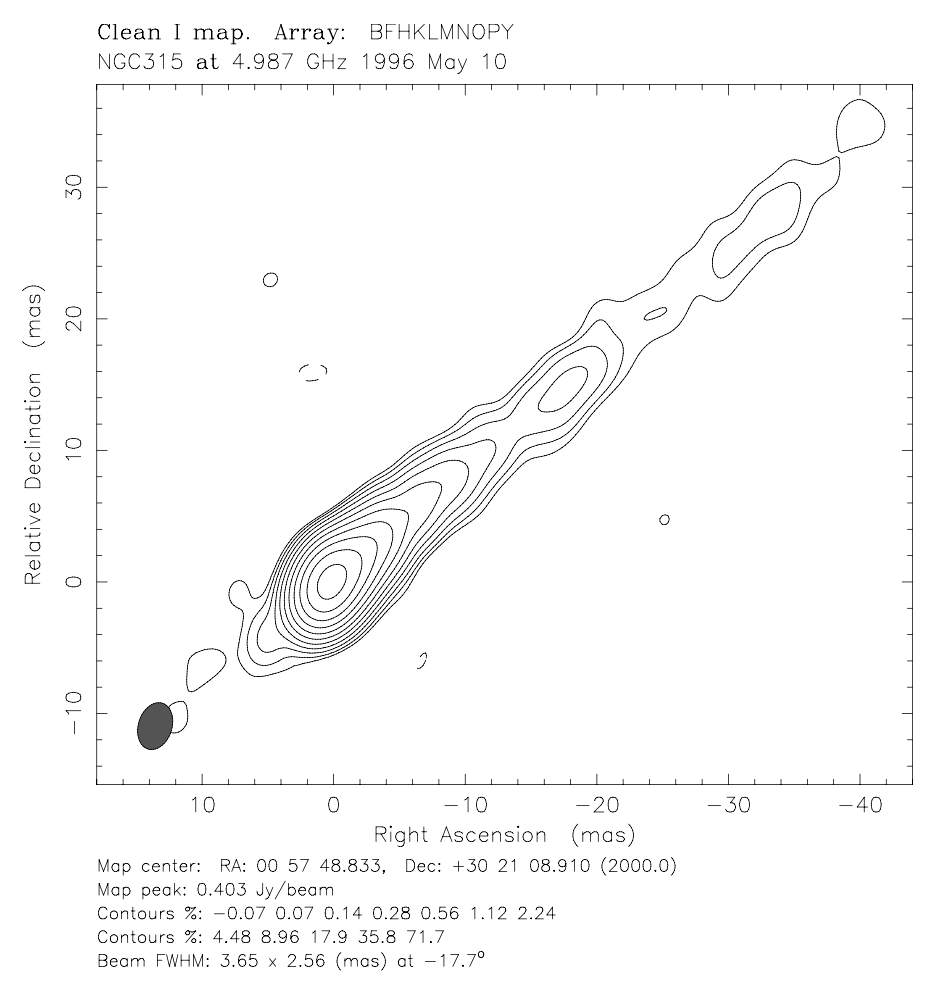}\par
    \includegraphics[width=6.3cm]{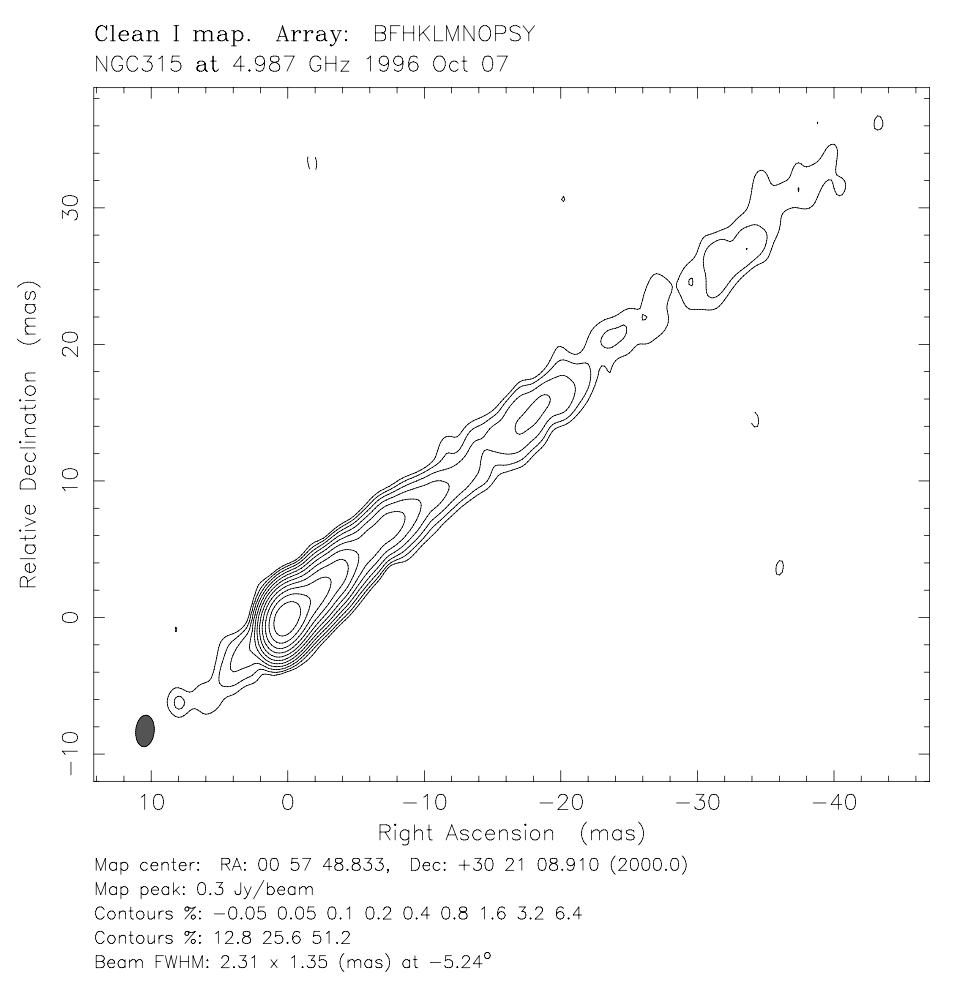}\par
\end{multicols}
\caption{VLBI images of NGC\,315 at 5 GHz.}
\label{fig:5GHz}
\end{figure*}

\begin{figure*}[h]
\centering
    \includegraphics[width=6.3cm]{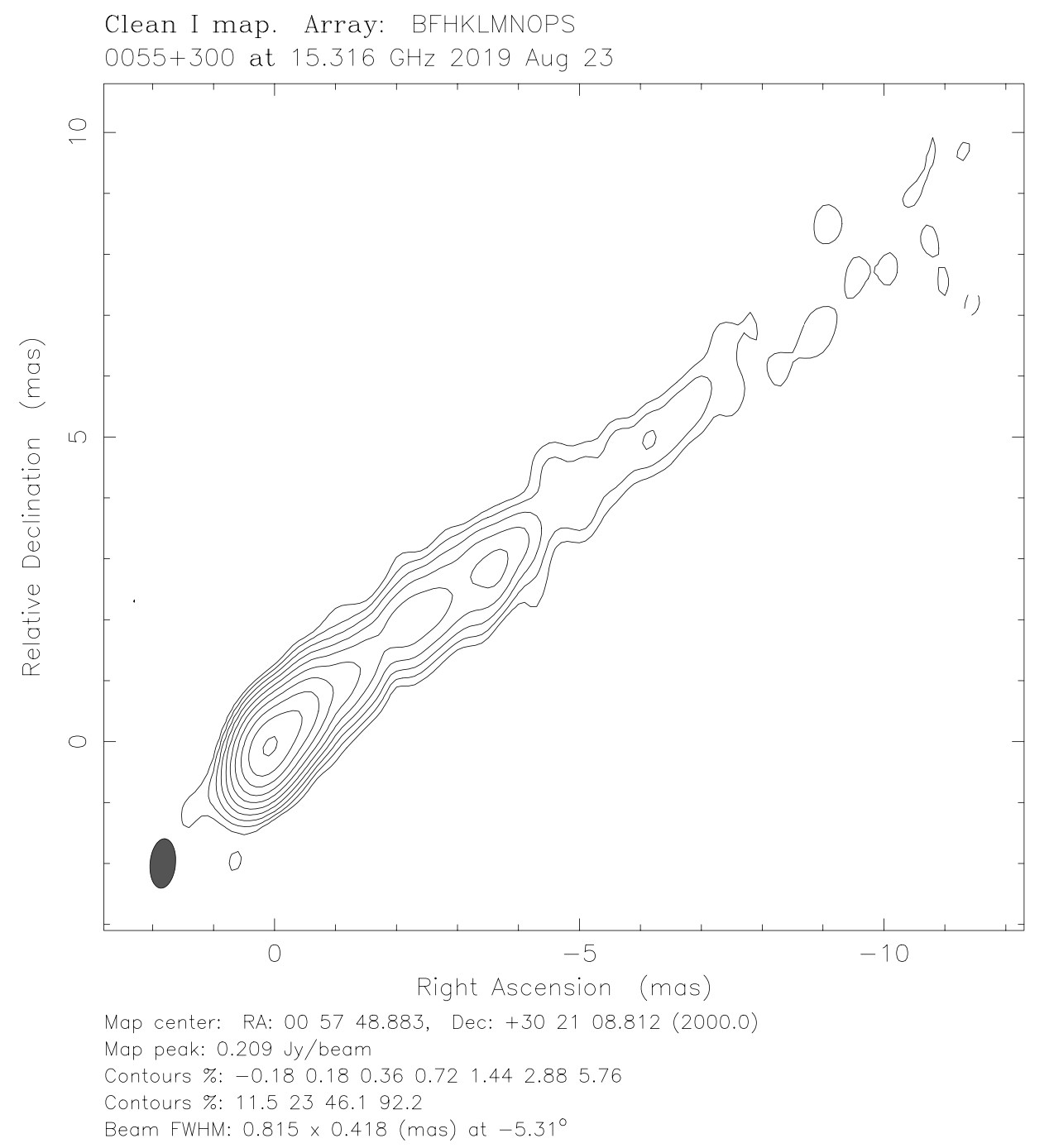}
  \caption{VLBI image of NGC\,315 at 15 GHz.}
  \label{fig:15GHzMOJAVE}
\end{figure*}

\begin{figure*}[h]
\centering
\begin{multicols}{2}
    \includegraphics[width=6.3cm]{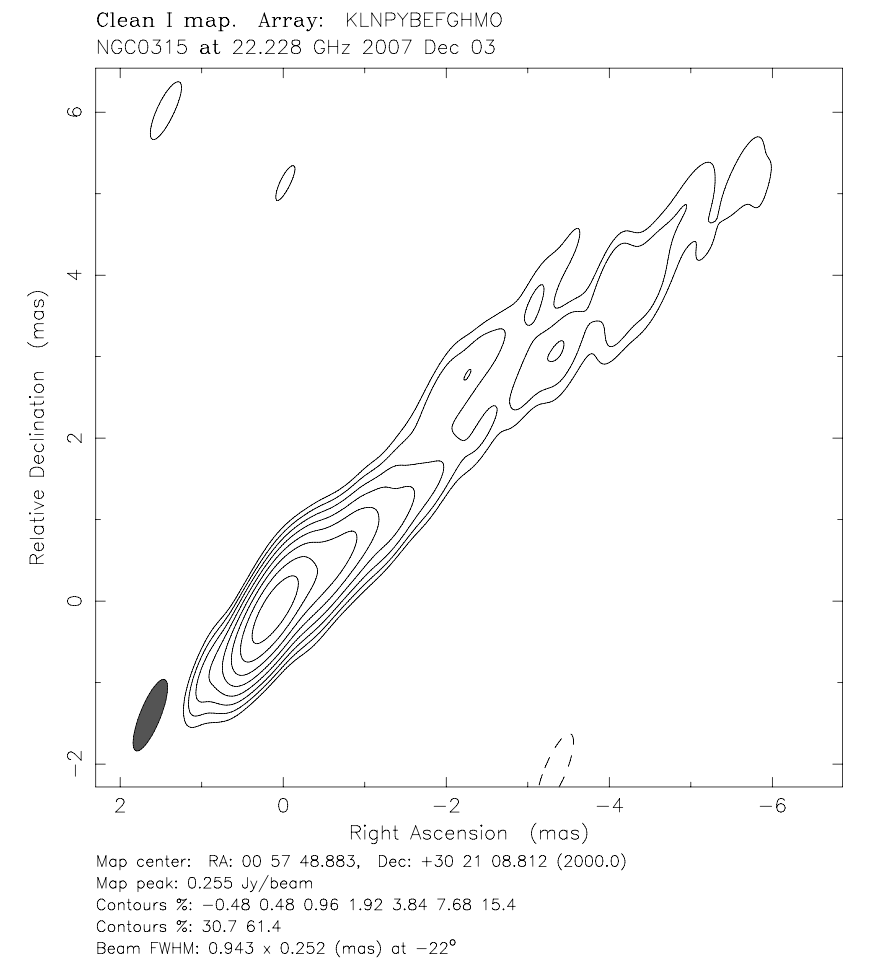}\par
    \includegraphics[width=6.6cm]{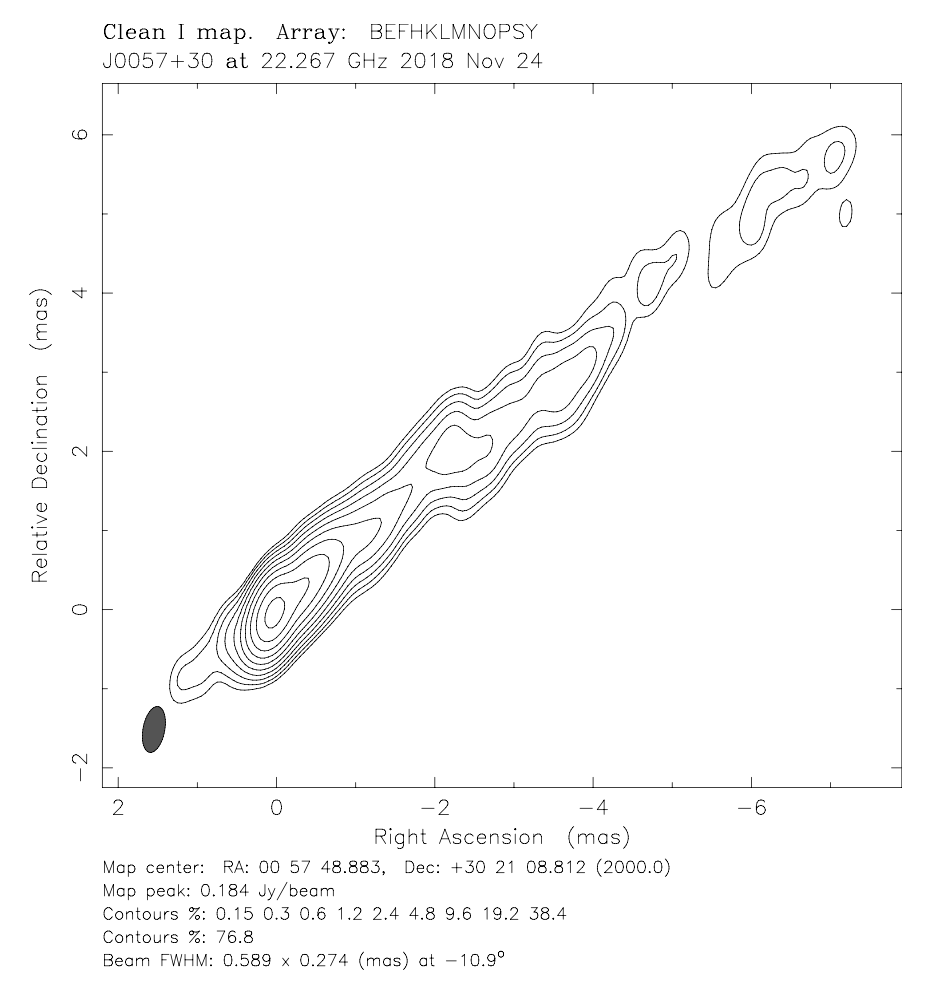}\par
\end{multicols}
\caption{VLBI images of NGC\,315 at 22 GHz.}
\label{fig:22GHz}
\end{figure*}

\begin{figure*}[h]
\centering
\begin{multicols}{2}
    \includegraphics[width=6.3cm]{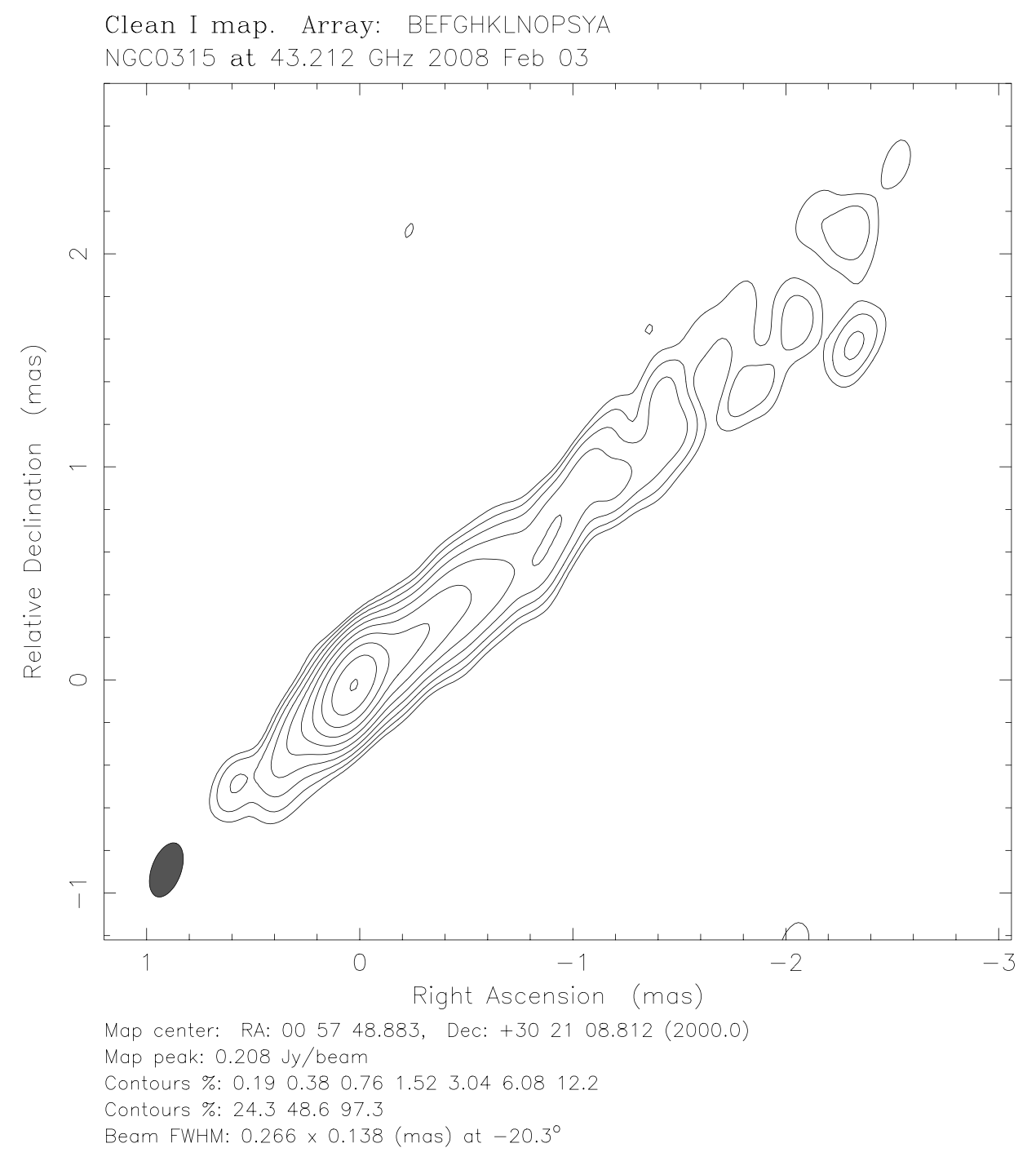}\par
    \includegraphics[width=7.2cm]{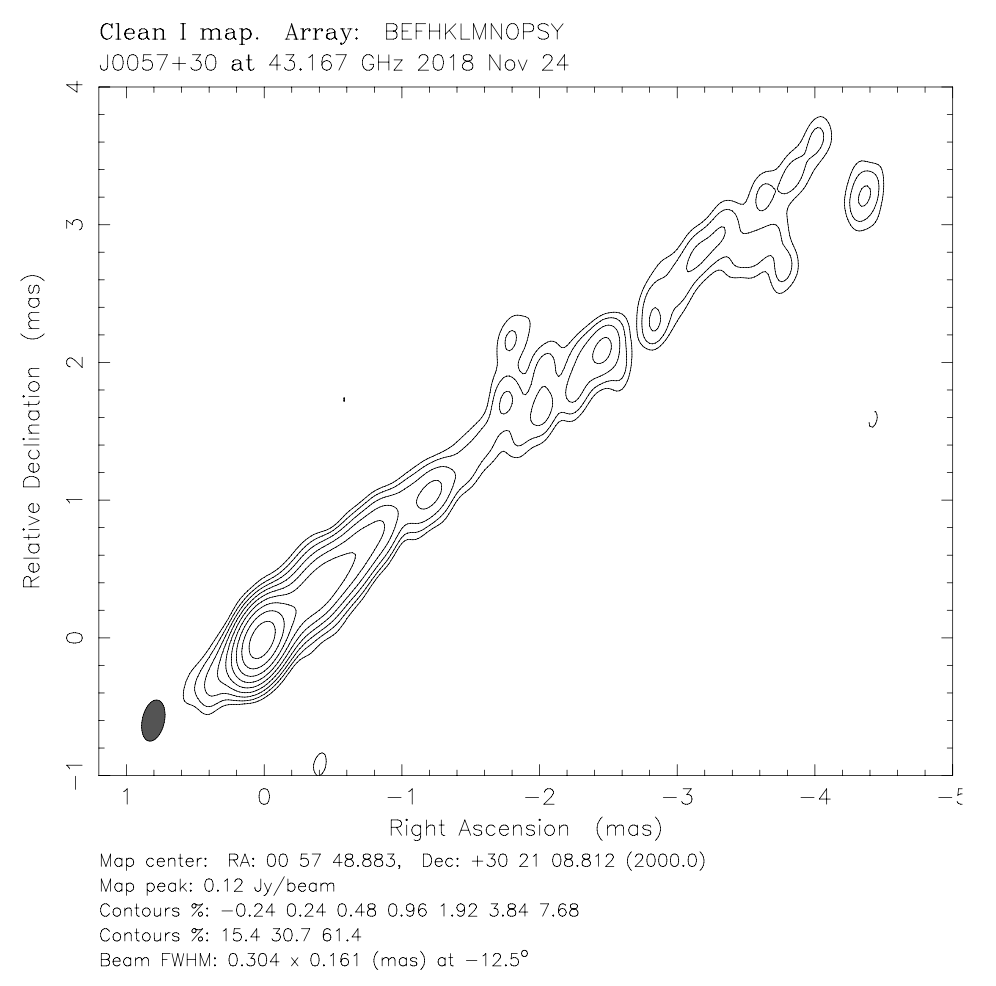}\par
\end{multicols}
\caption{VLBI images of NGC\,315 at 43 GHz.}
\label{fig:43GHz}
\end{figure*}

\end{document}